\documentstyle[12pt]{article}

\newcommand{\PSbox}[3]{\mbox{\rule{0in}{#3}\includegraphics{#1}\hspace{#2}}}
\newcommand{\be}{\begin{equation}}
\newcommand{\ee}{\end{equation}}
\newcommand{\bea}{\begin{eqnarray}}
\newcommand{\eea}{\end{eqnarray}}
\newcommand{\om}{\omega}
\newcommand{\Om}{\Omega}

\newcommand{\im}{\dot{\imath}}

\newcommand{\al}{\alpha}
\newcommand{\ga}{\gamma}
\newcommand{\Ga}{\Gamma}
\newcommand{\de}{\delta}
\newcommand{\De}{\Delta}
\newcommand{\ep}{\epsilon}
\newcommand{\E}{{\cal E}}
\newcommand{\si}{\sigma}

\newcommand{\la}{\lambda}
\newcommand{\half}{\frac{1}{2}}
\begin{document}
\thispagestyle{empty}
\begin{center}
{\Large\bf Plasma Resonance in Layered Normal Metals and Superconductors.}\\
\vspace{1cm}
{\large S. V. Pokrovsky}\footnote{To whom all correspondence should be
addressed.}\\
Department of Physics, MIT, Cambridge, MA 02139\\
    and\\
{\large V. L. Pokrovsky}\\
Department of Physics, Texas A\& M University,\\
College Station, TX 77843-4242
\footnote{Present address.}\\
and\\
Landau Institute for Theoretical Physics, \\
Kosygin str. 2, Moscow 117940, Russia.
\end{center}

\begin{abstract}
A microscopic theory of the plasma resonance in layered metals is presented.
It is shown that electron-impurity scattering can suppress the plasma resonance
in the normal state and sharpen it in the superconducting state.
\end{abstract}

\section{Introduction.}
\label{introduction}

The plasma resonance can be observed as a steep drop of  reflectivity
when the frequency of an electromagnetic wave goes above
the threshold plasma frequency $\Om$. The latter is determined
in an isotropic plasma or a metal by the well-known equation:
\be
	\Om^2=\frac{4\pi n\,e^2}{\E m}
\label{plasma-frequency}
\ee
where $n$ is the
density of the carriers (electrons), $m$ is their effective
mass and $\E$ denotes high frequency dielectric constant due to
the inner atomic shells.

If a metal can become superconducting, its London
penetration depth $\la$ at zero temperature is related to the
plasma frequency via a simple relationship:
\be
	\Om\,=\,\frac{c}{\la \sqrt{\E}}
\label{frq-depth}
\ee
where $c$
is the speed of light.
\par Experimental observation of the plasma resonance in
conventional metals is difficult because the plasma frequency
usually occurs in the far ultra-violet range, where inner photoeffect
processes are substantial. The situation changes drastically
for layered metals when the field polarization is
perpendicular to the layers.  Indeed, one can  naively  apply
eq. (\ref{plasma-frequency}) with a large effective mass $M$ corresponding to
the
tunneling motion across the layers.  However, we show that the
naive approach fails due to the openness of the Fermi-surface.
The correct answer is at least $m/M$ times less ($m$ being a
small mass for in-plane motion). Thus the plasma frequency
$\Om_z$ for $c$-axis polarization occurs in the far infrared or
even in the millimeter range.

Polarized reflection measurements from $Y Ba_2 Cu_3 O_7$ crystals
have been carried out by Koch et al \cite{Koch}. For the  radiation polarized
perpendicularly to the $CuO_2$ planes  a striking difference has been
discovered between the reflectivity spectra in the normal and in the
superconducting state.  Typical spectra  below the superconducting transition
($90 K$) show a sudden sharp drop from  unity to $0.4$  in the vicinity  of
$12  meV$ $(90 cm^{-1})$ at $T =  10 K$ and to $0.55$ in the vicinity  of
$8  meV$  $(60 cm^{-1})$ at $T =  65 K$ whereas at temperatures  above $T_c$
reflectivity just decreases smoothly from $0.8$ to $0.60-0.65$ in the whole
far infrared region $0-15 meV$.

A similar behavior of the far infrared $c$-polarized reflectivity from
the $La_{2-x}Sr_{x}CuO_4$  single crystals has been reported by Tamasaku
et al \cite{tamasaku}.
Plasma edge followed by a dip has been observed by them
in the superconducting phase at even lower frequencies
($25 cm^{-1}$ at $ x = 0.10$ and $ 50 cm^{-1}$ at $x = 0.16$).
This spectral feature being very sharp at low temperature becomes smeared
when the temperature is increased and eventually disappears in the vicinity
of the critical point.

The physical reason for this phenomenon can be readily  understood. In the
normal state a strong scattering of the electrons by impurities and phonons
suppresses the plasma oscillations.  However, in the superconducting state
there exists a superfluid of  Cooper pairs which are not scattered,  and thus
maintain the coherence of the plasma wave.

The idea of abnormal transparency of layered
superconductors has been advocated by T.Mishonov \cite{todor}.
He pursued a phenomenological approach based on the two-fluid model
of superconductivity (Gorter-Kazimir model).  In this model the dielectric
function
is postulated in the following form:
\be
	\E \left( \om \right) = \E \left( 1 - \frac{\Om_s^2}{\om (\om + \im 0^+)}
	-  \frac{\Om_n^2}{\om (\om + \im /\tau)} \right)
\label{two-fluid}
\ee
The first term in the r.h.s. of this equation describes the contribution of
the
valence electron bands and high frequency optic oscillations. The second term
is
due to the Cooper pairs and the last one is contributed by the Fermi
excitations. The superconducting and the normal plasma frequencies,
$\Om_s$ and $\Om_n$, are related to the corresponding BCS densities
$n_s$ and $n_n$ in the same way as in eq. (\ref{plasma-frequency}).
The absorption is characterized by some temperature-dependent
attenuation coefficient $1/\tau$. This approach has been first employed
for the analysis of the optical properties of the high-$T_c$ superconductors
by de Marel {\it et al}  \cite{Marel}.
A recent theoretical work by Tachiki et al \cite{tachiki} goes along
the same lines.

The purpose of the present work is to present a microscopic theory of
the plasma resonance in the normal and the superconducting states of
layered metals. The anomalously low value of the plasma frequency for
the $c$-polarized radiation is a straightforward outcome of this theory.
Besides, it provides an interesting and important insight into the role
played by elastic scattering of electrons by impurities. The scattering
puts down the reflection of electromagnetic waves by
destroying the coherence of plasma oscillations in the normal state
once the scattering rate is comparable with the plasma frequency.
In the superconducting state, however, the frequent scattering prevents
the normal carriers to be involved into the collective motion of
the superconducting Cooper pairs.
Therefore, plasma oscillations of the charged superfluid being
virtually independent maintain the reflection below plasma edge.

The paper is organized as follows.
We introduce the underlying model in Section \ref{sec:model}.
The reflectivity in the normal state is evaluated in Section
\ref{sec:norm-refl} using
the kinetic equation. This approach is insufficient to treat the same
problem in the superconducting state. Not only the quasi-particle occupation
numbers, but also the Cooper pair wave function varies in the electromagnetic
wave (see \cite{PSav},\cite{eli}), so that no local dynamics exists for
superconductors. An outline of the field-theoretical calculations
accounting most naturally  for the above-mentioned variations and for the
random fields of impurities is drawn in Section \ref{sec:field-theory}.
The results of these calculations are represented and discussed in
Section \ref{sec:super-resonance}. We return to the
discussion of advantages and limitations of our model and
summarize its physical implications in the last Section \ref{sec:conclusions}.
The analytic continuation of the electromagnetic response function  is
carried out in \ref{app:analyt-cont}.  The electromagnetic response
for the frequencies above the threshold $2 \, \De$ is studied in
\ref{app:above-threshold}.
\par
Preliminary version of this work has been published in the
Proceedings of SPIE conference.

\section{The model.}
\label{sec:model}

     Most layered conductors have a rather
complicated crystal structure including, in particular, several
conducting layers ($ CuO_2$) per elementary cell. These layers
are usually coupled much more strongly than those belonging to
different cells. In this article a simplified model with a single
layer per elementary cell is employed. To make the model even
simpler, the electron dispersion relation  is taken to be totally
isotropic in the $ab$-plane. Although the spectrum of the real
compound possesses only orthorhombic symmetry, the
anisotropy in the $ab$-plane is not very pronounced. In other
cases the symmetry may be tetragonal. That gives us hope that the
isotropic model provides a reasonable approximation.

     In our model the electrons are assumed to be scattered
predominantly by the impurities. The collisions are assumed to
be elastic, and since the overlapping  of the electronic wave
functions even between the neighboring planes is small, only
in-plane scattering is taken into account. In
principle, two competing types of processes, that of in-plane and
inter-plane scattering  with different collision times $\tau$
and $\tau^{\prime}$, respectively, should be considered
independently. Note that the
inter-plane collision time $\tau^{\prime}$ must be compared to
the effective time of inter-plane quantum tunnel hopping
suppressed by the in-plane scattering \cite{Varlamov}. This
time is equal to $(\ga^2 \tau)^{-1}$ where $\ga$ is an
inter-plane hopping amplitude. We assume the
strong inequality $\tau^{\prime} \gg (\ga^2 \tau)^{-1}$ to be
valid, allowing us to disregard the inter-plane scattering.
Our  theory relies
on the assumption that the metallic properties of the $CuO$ planes are
not spoiled by scattering,
i.e. $\ep_F\tau \gg 1$ ($\ep_F$ stands for the Fermi energy).
On the other hand, the dimensionless parameter $\gamma \tau$ measuring
the relative value of the inter-plane tunneling with respect to the
in-plane scattering
may vary in a broad  range.  The role of inelastic processes and other more
realistic modifications to the theory will be considered
elsewhere.

To summarize, our model is that of non-interacting
massive electrons propagating freely in the {\sl ab}-plane and
hopping to the neighboring planes. Their motion in the ideal
crystal is governed by the following dispersion law
\be
     \epsilon (\vec{p}, p_z) = p^2/2m
     + \gamma \left( 1 - \cos{(p_z d/ \hbar)} \right)
\label{dispersion}
\ee
where $\vec{p} = (p_x, p_y)$ is an in-plane component of momentum
and $p_z$ is a quasi-momentum component parallel to $c$-axis.
The effective mass and the inter-layer spacing are denoted by $m$
and $d$ respectively. Finally $\gamma$ is the hopping amplitude
introduced earlier.

 The following obvious relationship between the hopping
amplitude $\gamma$ and the effective mass in $c$-direction $M$

\be
     \gamma =  \hbar^2 /M d^2
\label{hop-amplitude-mass}
\ee
will be employed later.

As far as the superconducting properties  are concerned only the simplest
ground state characterized by an isotropic and homogeneous distribution of
the order parameter is considered.  The BCS approximation is adopted.
Although the  potential  drawbacks of  the  a weak coupling theory
for the quantative description of  high-$T_c$  superconductors are clearly
understood,
we believe that the basic features of the plasma resonance phenomenon are
revealed in the framework  of this approach.

\section{Reflectivity in the normal metal.}
\label{sec:norm-refl}
We now evaluate the reflectivity for the incident wave
propagating along $x$-axis, the surface of the layered metal
coinciding with the $yz$-plane, and electric field being normal to
the layers, i.e. directed along $z$-axis (c-axis). The current
$
	J_z (\om, k_x)
$
induced by the electric field
$
	E_z (\om, k_x)
$
in the skin layer is also directed
along the $z$-axis:
\be
	J_z (\om, k_x) = \si_{zz}(\om) E_z(\om, k_x)
\label{conductivity}
\ee
where $\om$ and $k_x$ are the frequency and the wave number of
the incident wave, respectively.
The conductivity $\si_{zz}$ depends in general both on wave number
and frequency
\be
	\sigma_{zz}(\omega,  k_x) = -\, 2 \im e^2
	\int \frac{\partial {n^{(0)}}}{\partial {\epsilon}}
	\frac{ v_z^2(\vec{p}) d^3 p}{\om + \im / \tau - v_x k_x}
\label{sigma-zz}
\ee
where $n^{(0)}(\ep)$ is the Fermi occupation number.
Neglecting space dispersion ($k v_F$ compared to
$
	\om + \im / \tau
$ )
one obtains a kind of Drude-Lorenz formula

\be
	\si_{zz}(\om) =
	\frac{\im e^2 \ga^2 m d}{2\pi\hbar^4 \left( \om + \im / \tau
		\right)}
\label{Drude}
\ee
Let us note that $\tau$ in the above expression is a true
collision time in contrast with the formula for the
homogeneous metals where it is substituted by a transport
time\footnote{The reason for this difference is the cancellation of
the "arrival" term in the collision integral due to the
independence of the scattering amplitudes on $p_z$.
}.

Henceforth we will omit tensor and vector indices of the conductivity
tensor $\sigma_{zz}$, electric
field $E_z$, magnetic field $B_y$ and wave vector and $k_x$
used in the present work and denote them simply $\sigma$, $E$,
$B$ and $k$, respectively.

The dielectric permeability of the metal is related to the conductivity
as
\be
	\E(\om) = \E + \frac{4 \pi \im}{\om} \si(\om)
	= \E \left( 1 - \frac{\Om_z^2}{\om (\om + \im / \tau)}
	\right)
\label{dielectric-constant}
\ee
where $\E$ denotes the high-frequency dielectric constant (a
contribution due to the valence electrons and high-frequency
optical modes), plasma frequency is defined as
\be
	\Om_z^2 = \frac{2 e^2 m}{\E M^2 d^3}
\label{plasma-frequency-z1}
\ee
The dielectric permeability,
in addition to being a function of $\omega$, depends on
two additional parameters
$
	\beta= v_F/c
$
and
$
	\Omega_z \tau
$.
Both of them are intrinsic characteristics of the metal. Although
$\beta$ is always small,
$
	\Omega_z \tau
$
may vary in different compounds.
We assume that it cannot be too large, obeying the following inequality
\be
	\beta \Omega_z \tau \ll 1
\label{small}
\ee
If (\ref{small}) is satisfied, the spatial dispersion of the
conductivity $\sigma$ may be neglected, as we have done
already. Indeed, one can verify that a strong inequality
$
	 k v_F \ll \mid \om + \im / \tau  \mid
$
is valid where
\be
	k =  \frac{\om}{c}  \sqrt{\E(\om)}
\label{penetration}
\ee
provided (\ref{small}) holds.

The reflectivity is related to the dielectric constant by the equation
\be
	R(\omega) =
	\left| \frac{\sqrt{\E(\om)} - 1}{\sqrt{\E(\om)} + 1} \right|^2
\label{reflectivity-def}
\ee
Several typical spectra of the reflectivity are
shown in Fig.1.  The steep drop of reflectivity appearing in the
vicinity of the plasma frequency $\Om_z $ when the latter is large
with respect to the scattering rate $1/\tau$ is usually followed by a dip
\footnote{We do not consider $\E \leq 1$. }.
The bigger is the high-frequency  dielectric constant $\E$,  the deeper is
the minimum.
When the impurity elastic scattering  rate $1/\tau$ is significant,
with respect to $\Om_z $ , the corresponding
reflectivity spectrum becomes featureless.

Let us emphasize two peculiarities of the interaction between the
electromagnetic field and layered metals. First, the plasma
frequency is surprisingly small. In fact, eq.
(\ref{plasma-frequency-z1}) for
$       \Omega_z^2 $
may be rewritten as
\be     \Omega_z^2 =    \frac{ 4 \pi N e^2}{\E M}
	   \frac{m}{M}     \frac{\hbar^2}{(p_{F}d)^2}
\label{plasma-frequency-z2} \ee
where $a$ is an in-plane
lattice spacing. The above expression for the plasma frequency
contains two small factors, $ m/M $
and $ (\hbar/p_{F}d)^2 $, reducing the conventional values by two or
three orders. Therefore, it is unlikely that
$
	\Omega_z \tau  \gg 1
$
in experiments on the copper oxides, at least until single
crystals of very high quality become available.  It looks even
less likely that the criterion (\ref{small}) can be violated.
Hence the Drude expression (\ref{Drude}) may serve as a good
approximation at any frequency. Recall that the
additional small factors appear as a consequence of the
openness of the Fermi-surface.
\par The second peculiarity concerns
the absence of the so-called anomalous skin-effect for
$z$-polarized radiation. In this anomalous regime, a leading
role is played by a strong spatial dispersion. The latter might
become essential once the mean free path of the electrons
exceeds the skin-layer depth. However, in this case the
inequality $ \beta \Om \tau \ll 1 $ would be violated, whereas
it is definitely satisfied in the experiments with
high-$T_c$ superconductors.

	To conclude this section, we argue that all our
results are robust with respect to the replacement of the quadratic
dispersion
$
	\epsilon(\vec{p}) = p^2/2m
$
by some arbitrary anisotropic dispersion
$
	\epsilon(\vec{p})
$,
provided the Fermi line (in the $ab$-plane) is sufficiently
smooth. Indeed, the essential relationship which allows to
neglect spatial dispersion is the inequality
$
	k v_F \tau \ll 1,
$
which is equivalent to the condition (\ref{small}). The
integration along the Fermi-line in (\ref{sigma-zz}) gives basically
the same formula  as for the quadratic dispersion. Only the
plasma frequency is redefined:
$$
	\Omega_z^2 = 2 \pi e^2 \gamma^2 d^2 N(\epsilon_F) / (\hbar^2 \E)
$$
where $N(\epsilon)$ is the density of states per unit volume.

\section{Field theory approach.}
\label{sec:field-theory}
For the reason given in the Introduction, we apply field
theory methods to the calculation of the dielectric permeability of layered
superconductors. If we set the gap in the spectrum, $\Delta$, to
zero, we get the result for the normal state. This justification of
the kinetic equation by more rigorous field theory methods is useful
since the validity of the kinetic equation is not obvious when
$\ga\tau<1$.

    The electromagnetic response tensor $Q_{\alpha \beta}$
relates the average current $\vec{J}$ to the vector-potential $\vec{A}$
\be
     J_{\alpha}(\omega, \vec{r}) =
     - \int Q_{\alpha \beta}(\omega, \vec{r}, \vec{r^{\prime}} )
	  A_{\beta}(\omega, \vec{r^{\prime}}) d^3 r^{\prime}
\label{response}
\ee
The quantity $Q_{\alpha \beta}(\omega, \vec{r}, \vec{r^{\prime}} )$ is
usually called the Pippard kernel.
     According to the Kubo formula, the response function (Pippard kernel)
$Q_{\alpha \beta}(\omega, \vec{r}, \vec{r^{\prime}} ) $
is equal to the Fourier transform of the retarded Green function
of two current densities:
\be
     Q_{\alpha \beta}(\omega, \vec{r}, \vec{r^{\prime}} ) =
     \frac{\im T}{c} \int_{0}^{\infty} dt e^{\im \omega t}
     \langle [J_{\alpha}( t, \vec{r}) , J_{\beta}(0,
\vec{r^{\prime}})] \rangle .
\label{Kubo}
\ee
where $T$ is the temperature of the system and the angular brackets
denote the averaging  over the Gibbs ensemble.
     Following  the conventional routine, we first compute the
auxiliary Matsubara response
\be
     Q_{\alpha \beta}^M (\omega_m, \vec{r},\vec{r^{\prime}} ) =
     \frac{\im T}{c} \int_{0}^{1/T} d \tau e^{\im  \omega_m \tau}
     \langle
	 J_{\alpha}( \tau, \vec{r}) J_{\beta}(0, \vec{r^{\prime}})
	\rangle
\label{Matsubara}
\ee
where $\omega_m = 2 \pi m T$.

     The retarded response (\ref{Kubo}) can be found via analytic
continuation of the Matsubara response (\ref{Matsubara}) from
the discrete points
$\omega = \im \omega_m  $
in the upper half-plane of the complex variable $\omega$, using the
well-known relation
\be
	Q_{\al\beta}^R(\im\om_m, \vec{r},\vec{r^{\prime}}) =
	Q_{\al\beta}^{M}(\om_m,, \vec{r},\vec{r^{\prime}})
\label{MRcon}
\ee
     If elastic impurity scattering only is taken into
account, the Matsubara Green function of currents in
(\ref{Matsubara}) may be expressed in terms of two-point
electronic Green functions. We prefer to write down the
corresponding relation in momentum space rather than in the
coordinate space; this will prove convenient for
subsequent averaging over impurity configurations:
\bea
     Q_{\alpha \beta}^M (\omega_m, \vec{k},
     \vec{k^{\prime}} ) =
    -
    \frac{e^2}{c}\int \frac{d^3 P}{(2 \pi \hbar)^3}
    \frac{\partial^2\xi(\vec{P})}{\partial P_{\al} \partial P_{\beta}}
    \, T \, \sum_{\nu}G^{M}(\eta_{\nu},\vec{P}_+ ,\vec{P}_-)
    -
    \nonumber\\
    \frac{\im e^2}{c} \int\, \frac{d^3p}{(2 \pi \hbar)^3} \,
    \frac{d^3q}{(2 \pi \hbar)^3} \,
    v_{\alpha}(\vec{p}) \,  v_{\beta}(\vec{q}) \,
    \Ga(\om_{m},\vec{p}_{+},\vec{p}_{-},\vec{q}_{+}\vec{q}_{-})
\label{Wick}
\eea
where $\eta_{\nu}=(2\nu +1)\pi T$, and
$
	\vec{p}_{\pm} = \vec{p} \pm \vec{k}/2,
	\vec{q}_{\pm} = \vec{q} \pm\vec{k^{\prime}}/2,
	\vec{P}_{\pm} = \vec{P} \pm (\vec{k}-\vec{k^{\prime}})/2  $,
and the group velocity
$
     v_{\alpha}(p) = \partial \xi / \partial p_{\alpha}
$.
Finally, the vertex $\Ga$, defined as
\bea
	\Ga(\om_n,\vec{p}_+,\vec{p}_-,\vec{q}_+,\vec{q}_-) =
	     T \sum_{\nu}
	G^M(\om_n + \eta_{\nu}, \vec{p}_+,\vec{q}_+)
	G^M(\eta_{\nu}, \vec{p}_-, \vec{q}_-)
	+
	\nonumber\\
	T \, \sum_{\nu}
	\bar{F}^M(\om_n + \eta_{\nu}, \vec{p}_+,\vec{q}_+)
	F^M(\eta_{\nu}, \vec{p}_-, \vec{q}_-)
\label{vertex}
\eea
depends on the four momenta $\vec{p}_{\pm}, \vec{q}_{\pm}$.
Let us emphasize that the electronic Green functions
are the exact Green functions for the non-interacting electrons
moving in the potential of some arbitrary but fixed
impurity configuration, and obeying the well-known  Dyson
integral equation \cite{AGD}. Averaging over the electronic Fermi
distribution has already been performed in (\ref{Wick}), but
averaging over the random ensemble of impurities
is yet to be done using the standard diagrammatic technique
\cite{AGD}.
After this second averaging, translational invariance is
restored, and the averaged Green functions depend on
twice as few variables:
\bea
	G^M(\eta_{\nu}, \vec{p},\vec{q})
	 \, =\, \de(\vec{p}-\vec{q})
		G^M(\eta, \xi(\vec{p}))
\nonumber\\
	F^M(\eta_{\nu}, \vec{p},\vec{q})
	\, =\, \de(\vec{p}+\vec{q})
		F^M(\eta, \xi(\vec{p}))
\label{after-averaging}
\eea

These reduced  Green functions were calculated by Klemm {\it  et al}
\cite{Beasley}, by a natural modification of the method of
averaging originally proposed by Abrikosov and Gor'kov
\cite{AGD} for isotropic superconductors:

\bea
	G^M(\eta, \xi) = - \frac{
	\im \eta \tilde{\ep}(\eta) / \ep(\eta)
	 +
	\xi}{ \tilde{\ep}(\eta)^2 + \xi^2}
\nonumber\\
	F^M(\eta, \xi) = \frac{\Delta \tilde{\ep}(\eta) / \ep(\eta)
		}{\tilde{\ep}(\eta)^2 + \xi^2}
\label{averaged-green-functions}
\eea
Here
\be
	\ep(\eta) = \sqrt{\eta^2 + \De^2}
	\hspace{0.5in}
	\tilde{\ep}(\eta) =
	\ep(\eta) + 1 /(2\tau)
\label{xi}
\ee

The averaging of the vertex (\ref{vertex}) normally involves a
solution of the Bethe-Salpeter equation or, more
accurately, a system of equations for different vertices. For
the dynamic processes  the number of components increases since
the vertex depends on two independent frequencies and can be
retarded or advanced with respect to each of them
\cite{keldysh,AGD}. Fortunately, these difficulties can be
avoided for the component $Q_{zz}$ of the response tensor. The
calculations in this case are simplified due to the
independence of the amplitudes of the electron scattering by
the impurities on the $z$-component of the electronic momentum.
In the framework of this approximation for the Bethe-Salpiter
diagrams, integration over the $z$-component of  the momentum
in each of the vertices can be separated and involves the
following integrals:
$$
	\int_{-\pi/d}^{\pi/d} \frac{dp_z}{(2 \pi)} v_z(p_z)
	G^M(\eta,\xi_+)         G^M(\eta,\xi_-)
$$
where $\xi_{\pm} = \xi \pm (\vec{k}, \vec{v})/2 $. These
vanish \footnote{This nullification corresponds to the
vanishing of the "arrival" term in the collision integral of
kinetic equation.} because $v_z$ is the total derivative of a
periodic function of $p_z$. According to the above arguments
the component of the response tensor $Q_{zz}$ is given by:
\be
     Q_{zz}^M (\omega_n, \vec{k}) =
	- \frac{ \im e^2}{c} \int
     \frac{d^3p}{(2 \pi)^3}
     v_{z}^2(p_z)
	\Ga^M(\om_n,\xi_+,\xi_-)
\label{Q-matsubara-zz}
\ee
where
\bea
	\Ga^M(\om_n,\xi_+,\xi_-) \, =\,
	T \sum_{\nu} \left[
	G^M( \om_n + \eta_{\nu}, \xi_+)
	G^M( \eta_{\nu}, \xi_-) \right.
	\nonumber\\
	+\,
	\left.
	\bar{F}^M( \om_n + \eta_{\nu}, \xi_+)
	F^M( \eta_{\nu}, \xi_-)
	\right]
\label{pseudo-vertex}
\eea
To obtain the Pippard kernel, one must perform the analytic
continuation of
the quantity $\Ga$ and then integrate it over $\xi$.
The details of  this procedure may be found in the Appendix A.

\section{Plasma resonance in a superconductor.}
\label{sec:super-resonance}
For the specific problem under consideration we neglect
spatial dispersion, i.e. the dependence of $Q_{zz}$ on the
wave-vector $\vec{k}$.
For the dielectric permeability we use eqs.
(\ref{dielectric-constant}) with the following
modification:
\be
	\E(\om) = \E - \frac{4 \pi c}{\om^2} Q_{zz}(\om)
\label{super-dielectric-const}
\ee
In the superconducting state the
penetration depth does not increase to infinity at zero
frequency due to the Meissner effect, but it still has its
maximum at zero frequency. We have argued in Section 3 that it
is much bigger than the mean free path at high frequencies.
This condition is satisfied without  doubt at low frequencies.
The following calculations are similar to the treatment of the
high-frequency response of  superconductors by Bardeen
and Mattis \cite{BM} and by Abrikosov, Gor'kov and Khalatnikov
\cite{AGK} .  We modified the method of the latter authors
to account for the layered structure and the impurity
scattering.
\par
After the analytical continuation
in the \ref{app:analyt-cont} (see eq. (\ref{A-tanhtransform}))
of the vertex $\Ga^M(\om)$
in the eq. (\ref{Q-matsubara-zz}),
the integration over $\xi$ is performed explicitly in
the eq. (\ref{A-after-xi-integration}).
We express the Pippard kernel $Q_{zz}$
\be
Q_{zz}=\frac{e^2\ga^2 m\,d}{4\pi\, c \, \hbar^4} K^R (\om ,T)
\ee
via the dimensionless function $K^R (\om ,T)$.
The formulae for its real and imaginary part in the frequency range
$\om < 2 \De$ derived straightforwardly from
the eq. (\ref{A-after-xi-integration})
are presented below:
\bea
	{\cal R}e\, K^R (\om, T) \, =\, \int_{\De -\om}^{\De}
	\frac{\tanh(\eta_+/2T)d\eta}
	{\ep_+^2 + (\bar\varepsilon +1/\tau)^2}\,
	\left[\ep_{+} + \frac{\eta\eta_{+}+\De^2}{\bar\varepsilon \ep_+}
	\left(\bar\varepsilon + \frac{1}{\tau}\right) \right]
\nonumber\\
	+
	\frac{1}{2}\int_{\De}^{\infty} \left[
	\frac{\ep +\ep_+}{(\ep +\ep_+)^2+1/\tau^2}
	\left( 1-\frac{\eta\eta_{+} +\De^2}{\ep\ep_+}\right)
	\left( \tanh\frac{\eta_+}{2T} +
	\tanh\frac{\eta}{2T} \right) \right.
\nonumber\\
	+
	\left.
	\frac{\ep_+ - \ep}{(\ep_+ - \ep)^2+1/\tau^2}
	\left( 1+\frac{\eta\eta_{+} +\De^2}{\ep\ep_+}\right)
	\left( \tanh\frac{\eta_+}{2T} -
	\tanh\frac{\eta}{2T} \right) \right] d\eta
\label{ReK}
\eea
\bea
	{\cal I}m\, K^R (\om, T) \, =\,
	\frac{1}{2\tau} \int_{\De}^{\infty}  d\eta
	\left[ \left(1 -
	\frac{\eta\eta_{+} +\De^2}{\ep\ep_+}\right)
	\left((\ep_+ +\ep)^2+\frac{1}{\tau^2}\right)^{-1} \right.
\nonumber\\
	 -
	\left.
	\left(1+\frac{\eta\eta_{+} +\De^2}{\ep\ep_+}\right)
	\left((\ep_+ -\ep)^2+\frac{1}{\tau^2}\right)^{-1}\right]
	\left( \tanh\frac{\eta_+}{2T} - \tanh\frac{\eta}{2T}\right)
\label{ImK}
\eea
where the notations has been changed as follows
\be
	\eta_+ =\eta + \om \hspace{0.5cm}
	\ep = \sqrt{\eta^2 -\De^2} \hspace{0.5cm}
	\ep_+= \sqrt{\eta_+^2 -\De^2} \hspace{1cm}
	\bar\varepsilon = \sqrt{\De^2 -\eta^2}
\ee
Although the normal limit of the Pippard kernel at finite frequency
cannot be extracted from the above relations, we have verified
(see discussion on the page \pageref{pg:normal-limit}
following the eq. (\ref{A-imk})) that
\be
	\left.     K^R(\om) \right|_{\Delta = 0}
	=  \frac{\om}{(\om + \im / \tau)}
\label{eq:normal-limit}
\ee
confirming our earlier calculation based on kinetic equation.
\par  Let us analyze the frequency behavior of  the kernel  in several regimes
corresponding  to different values of temperature and  different
levels of scattering rate.
The main difference between the superconducting and the normal states is
that $Q_{zz}(\om)$  vanishes for $\om =0$ in the normal state,
but it takes a positive finite value in the superconducting state in
conformity with the London relation between
the current and the vector-potential. The London penetration depth
$\la_z$ is connected to the value of the kernel $Q_{zz}(0)$:
\be
	\la_z^{-2} = 4\pi Q_{zz}(0)/c
\label{LondonQ}
\ee
Two first terms in the eq. (\ref{ReK}) for
${\cal R}e\, K^R(\om)$
contribute to the static limit of the Pippard kernel:
\be
	K^R(0) = \pi \Delta \tau \tanh
	\frac{\Delta}{2T} - \De^2\int_{\Delta}^{\infty}
	\frac{d \eta \tanh(\eta/2T)}{\sqrt{\eta^2 - \Delta^2}
	\left[ \eta^2 - \Delta^2  + 1/4 \tau^2
	\right] }
\label{static-limit} \ee
This London factor vanishes as  $\Delta \longrightarrow 0$.

The Pippard kernel may be evaluated exactly in the clean limit :
\be
	\lim_{\tau  \longrightarrow \infty}  K^R(\om,T) = 1
\label{clean-kern}
\ee
Notice that  the limiting value doesn't depend on temperature.
This result is naturally associated with the suppression of dissipation.
The  Cherenkov absorption of photons is forbidden since the
Fermi  velocity is much less than the speed of light. On the other hand
breaking of Cooper pairs is kinematically permitted  but its rate is
proportional to the Bogolyubov coherence factor \cite{Rickayzen}
$
	\left( u_{\vec{p}+\vec{k}} v_{\vec{p}}
	- v_{\vec{p}+\vec{k}} u_{\vec{p}} \right)^2
$
where  $\vec{k}$ is the momentum of the photon.
For  the  normal skin effect ($\vec{k}=0$),
which we only consider,  this factor is equal to zero.
\par
Comparing the clean limit  to the static one
we come to conclusion that in a weak scattering regime
($\De \, \tau \gg 1$) and for the  temperature relatively
close to the transition point the real part of the
Pippard kernel grows rapidly from its static value
(\ref{static-limit}) to unity.
This onset becomes less pronounced at lower temperatures (Fig 2A).
The increase of scattering rate is accompanied by reduction of
the mean level of ${\cal R}e K(\om)$ and smothering of its variation
in the low frequency range (Fig 2B).
\par
At higher frequencies there is a plateau $K^R(\om,T) \approx 1$
ending in the vicinity of the threshold $\om = 2 \De$. This is the point
of a mild logarithmic singularity
\be
	{\cal R}e K^R(\om,T) = \frac{\pi \tau (\om - 2\De)}{2}
	\log\left(\frac{2\De-\om}{\De}\right)
	\tanh{\frac{\De}{2T}} +
	regular~ terms
\label{eq:log-sing}
\ee
where the derivative with respect to $\om$ becomes logarithmically large.
This singularity is closely related to the Cooper pair breaking above the
threshold (see discussion on the page \pageref{pg:log-sing}).
\par
Thus, by changing the temperature one can observe a large variation of the
plasma frequency when $2\De$ approaches the bare plasma frequency $\om_p$
but still is larger than the latter.
The attenuation can remain reasonably small to provide a sharp plasma
resonance since $\Om,\tau$ and $T_c$ are independent parameters.
\par
The observable plasma frequency at a given temperature
is defined as a root of equation
\be
	{\cal  R}e \E(\om_p) = 0
\label{plasma-frequency-equation}
\ee
In a sufficiently clean superconductor  ($\De \,\tau \gg 1$) and
for the moderate values of the bare plasma frequency $\Om$
(\ref{plasma-frequency-z2}) the observable plasma frequency $\om_p$
almost coincides with the bare one.  However, if
$\Om \, \tau \leq 1$ the observable plasma frequency $\om_p$
diminishes reaching its static limit at
$\om \ll 1/\tau$.  In more dirty samples the deviation of the
observable plasma frequency from the bare one is very significant
in the whole range $0 \leq \Om \leq 2\,\De$.
The diminution of temperature causes the augmentation of
the observable plasma frequency. Nevertheless, in a very dirty limit
($\De \, \tau \ll 1$) the observable plasma
frequency is by a factor $\sqrt{\De\,\tau}$ smaller than the bare one
even at zero temperature.

The behavior of the real part of the Pippard  kernel is illustrated in
Fig.2A ($\tau \ T_c  = 0.1$) and Fig.2B ($\tau \, T_c = 3.4$).  Two plots
are presented on both figures: dashed for $T/T_c = 0.3$
and dashed-dotted $T/T_c = 0.74$ respectively. Looking at the dirtier case
one can see that the observable plasma frequency is less than its static
limit $c/\la$ at the lower temperature and exceeds this limit at
the higher  temperature. It would be very instructive to study experimentally
the  temperature variation of the plasma frequency in more detail
and to compare it to the temperature dependence of the London penetration
depth in  $c$-direction $\la_z$. The latter  may be obtained independently
from the measurements of the radio-frequency impedance. Our theory
predicts extremely weak temperature
dependence of the observable plasma frequency in clean samples
whereas this dependence may be fairly remarkable in
samples with a moderate concentration of impurities.
\par
The imaginary part of the Pippard kernel is determined by the dissipation.
As it has been already mentioned, two types of photon absorption  are
substantial:
the Cherenkov absorption of a photon  by a quasiparticle (Landau damping) and
the absorption accompanied by the dissociation of a Cooper pair. The
former process is forbidden kinematically in the absence of scattering,
the latter is suppressed for the photon momentum $\vec{k} =0$.
At a finite scattering rate the momentum
is not  conserved any more, and both processes become possible.
The pair dissociation requires the photon frequency $\om >2\De$.
This range of frequencies is considered in
\ref{app:above-threshold}.
\par
The probability of the Cherenkov  absorption is proportional to the Fermi
occupation numbers of quasiparticles. These vanish exponentially at
$T\longrightarrow 0$.
The  asymptotic form of ${\cal I}m\, K^R$ for $ \De \gg  T $
and  $\om  \gg  T$ can be readily found from  eq.(\ref{ImK}):
\be
{\cal I}m K^R(\omega) =
	-\,e^{-\Delta/T}\,
	\sinh\left( \frac{\omega}{2\,T}\right)
	{\rm K}_0\left(\frac{\omega}{2\,T}\right)\,
	\frac{2\,\Delta\,\tau\,\sqrt{2\,\left( 2 + \omega/\Delta\right)}}
 {\left( 1 + 2\,\omega\,\Delta\,\tau^2\,
	 + \,\tau^2 \,\omega^2\ \right) }
\label{interpol-imagin-kernel}
\ee
where  ${\rm K}_0$ is a MacDonald function  of the zeroth order.
Examining it one can find that the dissipation depends substantially on the
scattering rate and in  a somewhat surprising way. Namely, it becomes
especially small provided  $\De \, \tau \ll 1$. Even as the temperature
approaches to the transition point and the number of the excitations
increases the dissipation may continue to
be small if the scattering rate is sufficiently high. The physical reason for
this phenomenon, as already mentioned in the Introduction, is the decoupling
of a strongly dampen excitation motion from the collective plasma
oscillations of the Cooper pairs. However, the asymptotic formula
(\ref{interpol-imagin-kernel}) does not apply in this case.
\par
Several plots of imaginary part of the Pippard kernel versus frequency are
shown in Fig.3. A minimum of
${\cal I}m K^R(\om )$ occurs at some frequency which
depends on $T_c$, $T $ and
$\tau$.
Several characteristic frequency scales for ${\cal I}m K^R(\om )$
can be easily seen in
eq. (\ref{interpol-imagin-kernel}). These scales  are
$T, \De, \tau^{-1}$ and
$(\De\tau^2 )^{-1}$. When $T$ is the smallest of the scales the minimum of
${\cal I}m K^R(\om )$ is located
at $\om\sim T$ and the minimal value of ${\cal I}m K^R(\om )$  may reach the
order of unity for the moderate temperatures in the clean case.
 We also see that ${\cal I}m K^R(\om )$ is presumably a linear function
of $\om$ at $\om\ll T$ and behaves as
$\om^{-1/2}$ at $\om\gg T$, but still $\om\leq \De$. One can readily find
numerous types of limiting behavior
of ${\cal I}m K^R(\om )$ depending on the relationship between
above-mentioned scales.
\par
Reflectivity spectra are shown in Fig.4.
The plots illustrate the behavior of the spectra with respect to
the temperature variation and the dependence on a dimensionless
parameter\footnote{We choose the transition temperature as a natural energy
scale.}
$\tau  T_c$, intrinsic for a specific sample.
The values of  two other sample-specific parameters:
high frequency dielectric constant and bare plasma frequency are chosen
to be $\E = 25.0$ and  $\Om_z /T_c = 1.2$
for each of the four plots. Each plot corresponding to some scattering rate
($\tau\,T_c =0.1;\,0.9;\,1.7;\,3.4$)  is combined from three subplots:
a solid curve for the normal metal and two for the superconducting state
at temperatures  $T/T_c = 0.3$ and $T/T_c = 0.74$.
At the largest value of the scattering
rate (Fig.4A) plasma edges followed by the "beaks" are clearly visible
on both subplots below $T_c$ while the normal subplot is completely
featureless. Subsequent plots show the evolution of the plasma feature
for the higher temperature. At the lowest scattering rate
$\tau \, T_c = 3.4$  (Fig.4D) the normal spectrum is still smeared whereas
the superconducting spectra for both temperatures:
$T=0.74\,T_c$ and $T = 0.3\,T_c$ are already close to the limiting
collisionless spectrum.

\section{Conclusions.}
\label{sec:conclusions}
We have shown that the plasma resonance can be suppressed in
the normal state of a layered metal due to a strong
electron-impurity scattering, but it should be restored in the
superconducting state since the superconducting part of the
electron liquid is not subject to scattering.
Moreover, below $T_c$ scattering sharpens the plasma
edge because it reduces the attenuation by removing Fermi-excitations
from the coherent wave motion.
\par
This feature is common for both the BCS and the two-fluid (Gorter-Kazimir)
models.
Indeed, the imaginary part of the dielectric function (\ref{two-fluid})
is small for the resonance frequency $\om_p = \Om_s$ provided a strong
inequality $\Om_n^2 \tau \ll \Om_s$ is satisfied. This condition is
obviously violated in the vicinity of the superconducting transition.
\par
Let us further discuss the relationship between the present theory and
the two-fluid model. Namely, we are going to address the question, to what
extent the phenomenological expression (\ref{two-fluid}) may approximate the
analytic structure of the real dielectric function which we believe is
correctly described by eqs.
(\ref{super-dielectric-const}-\ref{ImK}). This issue is important since the
Kramers-Kronig transformation  used routinely in the analysis of the
experimental reflectivity data is very sensitive to the singularities of the
analytic function involved.
\par
Let us first compare the behavior of the real part of the dielectric function
in a weak scattering regime: $\De \tau \gg 1$. In this case the BCS behavior
described in the previous section is  reasonably well reproduced in the
frameworks of the two-fluid model. In particular, the static limit of $\om^2
{\cal R}e \E (\om)$ is proportional to the density of superconducting Cooper
pairs.  The growth of the response real part due to the contribution of the
normal carriers saturates at the same level as in the BCS model when $\om \gg
1/\tau$. However, even in the clean limit the  shapes and (which may be more
important) the characteristic scales of the low frequency onset are different.
Indeed, the intermediate frequency scale $1/\De^2\,\tau$ arising naturally in
the BCS theory cannot be found in the Gorter-Kazimir model.
Finally, it is impossible to incorporate  the threshold phenomena in the
vicinity of $\om = 2 \De$, like a logarithmic singularity of ${\cal R}e
K^R(\om,T)$
described by  the eq. (\ref{eq:log-sing}), into the two-fluid plasma
electrodynamics.
\par
The augmentation of scattering worsens the agreement between the two  models.
In a strong coupling regime: $\De \tau \ll 1$ the two-fluid model predicts
an anomalously long plateau at the level of the static limit because the
normal electrons are overdampen and do contribute only to the imaginary part
of the Pippard kernel. The BCS strong coupling behavior of the
${\cal R}e K^R(\om)$ is different (see Fig. 2B). In fact, the two-fluid model
encounters
some principal difficulties  at small enough $\De\tau$.
It is not clear how the normal and superfluid
densities can be defined consistently when the non-conservation of the
momentum becomes significant.
\par
In the work \cite{tamasaku}
the clean limit of the two-fluid model has been considered. This approach
looks dubious because the fact that the resonance is suppressed in the normal
state combined with the experimental value of $\om_p\sim 2/3 \De$
imply that $\De\tau < 1$.
\par
As far as ${\cal I}m K^R (\om)$ is concerned one can find less common
features than the discrepancies between the two-fluid and the BCS models.
It is not surprising, since the imaginary part of the
Pippard kernel as a function of $\om$ has only one characteristic
scale, $\tau^{-1}$, whereas it depends on four different scales:
$\tau^{-1}, \De, T, (\De\tau^2)^{-1}$ in the present theory. In
particular, the position of the minimum of the function
${\cal I}m K^R (\om)$ in the two-fluid theory it is always $\om_{min}=
\tau^{-1}$. In the BCS calculations it depends strongly on the temperature.
At low temperature $\om_{min}\approx T$
and is always smaller than $\tau^{-1}$. A striking difference between the BCS
and the two-fluid dissipation in a strong scattering regime
($\tau T_c = 0.1$) is demonstrated in Fig. 5A.
However, even in the clean superconductor the dissipation calculated
in the frameworks of the BCS model is much weaker than its two-fluid
counterpart (Fig 5B). Of course the light absorption through the Cooper pair
breaking remains beyond the scope of the two-fluid model.
\par
Summarizing, the two-fluid model gives a reasonable, though in
some range of variables rather crude approximation for
${\cal R}e \E (\om)$
and fails to describe correctly ${\cal I}m \E(\om)$.
\par
Although many realistic features are lacking in the underlying model
 we believe that our main findings will survive in
its more refined and sophisticated versions. It is likely
that the in-plane anisotropy of the electronic dispersion
would change the macroscopic constants, but
not the shape of the plasma resonance.
It is not so clear, however, what would be the impact of the anisotropy
of the superconducting gap. One may conjecture
e.g. that if the d-wave pairing mechanism of superconductivity prevails
the attenuation at low temperatures would not be exponentially small.

In a more ambitious treatment, one may have to reexamine the
scattering model.
According to the experimental data \cite{static-conduct} the
resistance in the $c$-direction displays a semi-conducting
rather than a metallic behavior. Therefore, the
inter-plane motion might be better described  in terms of
hopping between the defects than between the planes.
We argue, however, that  at high frequencies
long-distance hopping is ineffective, whereas
short-range hopping is regulated by inter-plane tunneling.
Nevertheless a detailed treatment of these problems is highly
desirable.

Although the plasma resonance has been observed only in
the superconducting state so far, future developments in a technology
of single crystal preparation could drastically raise the
chances of its observation in the normal state. Such single
crystals could be utilized as 100\% polarizers in the far infra-red
region.

It would be very interesting to search for the plasma
resonance in the more pronounced layered superconductors, such as
Tl- and Bi-based oxides, where the edge could occur in
the microwave region. Especially promising would be epitaxially grown
multilayers, in which the transition temperature and the energy gap are
determined by the superconducting layers whereas the plasma frequency is
determined by the square of the inter-plane tunneling amplitude.

{\bf Acknowledgments.}
We are indebted to P.A. Lee for attracting our attention to the work
\cite{tamasaku} and to T. Mishonov and L.N. Bulayevskii for useful
discussions. Our special thank is due to I. Bozovic and W. Saslow for their
interest in this work and attentive reading of the manuscript.

\renewcommand{\thesection}{Appendix \Alph{section}}
\setcounter{section}{0}
\section{Analytical continuation.}
\label{app:analyt-cont}

\renewcommand{\thesection}{\Alph{section}}
\renewcommand{\theequation}{\thesection.\arabic{equation}}
\setcounter{equation}{0}

According to Section \ref{sec:field-theory} (eq. (\ref{MRcon})), to
calculate the Pippard kernel one must perform an analytic continuation of
the quantity $\Ga$ defined by eq. (\ref{vertex}) which is reproduced here
for visibility omitting the factor $-2 \im e^2/c$:
\bea
	\Ga^M(\om_n,\xi_+,\xi_-) &=&
	T \sum_{\nu}
	G^M( \om_n + \eta_{\nu}, \xi_+)
	G^M( \eta_{\nu}, \xi_-)
	\nonumber\\
	&+&
	T \sum_{\nu}
	\bar{F}^M( \om_n + \eta_{\nu}, \xi_+)
	F^M( \eta_{\nu}, \xi_-)
\label{A-pseudo-vertex}
\eea
from  a discrete set of points
$\om_n = \im 2n \pi T$ with $n \geq 0$ into the whole upper half-plane of the
complex variable $\om$.
We have restored the spatial dispersion omitted in the text,
since the analytic continuation can be performed in a more general context.
\par
Once analytic continuation is performed the integration over
$\xi$ is in turn.  Let us note, however, that
the integral over $\xi$ taken in the infinite limits and the sum over
$\nu$ together are divergent. Instead of introducing an
explicit cutoff in course of calculations one can deal with convergent
expressions employing a proper regularization procedure. To this end we
subtract from $\Ga (\om)$ the same quantity at
$
	\om = 0
$
and
$
	\Delta = 0
$.
The corresponding Green functions will be labeled by an additional
subscript $0$.
\par We start from the analytical continuation of the Green functions
calculated
by Klemm {\it et al }(\ref{averaged-green-functions},\ref{xi})
from a discrete set of arguments $\eta_{\nu} = \im \pi (2\nu +1)$. First, we
rewrite these functions in the following
form:
\bea
	G^M(\eta , \xi) = - \frac{\im \left( \eta + \eta /(\ep(\eta)\tau)
		\right) +\xi}
		{ \left( \ep(\eta) + 1/2 \tau \right)^2 + \xi^2}
\nonumber\\
	F^M(\eta , \xi) = \frac{\Delta \left( 1+(\ep(\eta)\tau)^{-1} \right)}
	{\left( \ep(\eta) + 1/2 \tau \right)^2 + \xi^2}
\label{A-averaged-green-functions}
\eea
Expanding the above expressions over the poles of the formal variable
$\ep(\eta)$ and analytically continuing the latter as a function of the
variable $\eta$ one obtains the
retarded Green functions in the following form:
\bea
	G^R(\eta , \xi) = \half \left(
	\frac{1 + \eta / \ep_R(\eta) }{ \ep_R(\eta) + \im /2 \tau - \xi}
	-
	\frac{1 - \eta / \ep_R(\eta) }{ \ep_R(\eta) + \im /2 \tau + \xi}
	\right)
\nonumber\\
	F^R(\eta , \xi) = \half \frac{\Delta}{\ep_R(\eta)}
		\left(
	\frac{1 }{ \ep_R(\eta) + \im /2 \tau - \xi}
	+
	\frac{1 }{ \ep_R(\eta) + \im /2 \tau + \xi}
	\right)
\label{A-retarded-green-functions}
\eea
where
\be
	\ep_R(\eta) = \sqrt{ (\eta+\im 0)^2 - \Delta^2}
\label{A-epsilon-retarded}
\ee
The branch of the square root is chosen to give positive values of
$
	\ep_R(\eta)
$
for
$
	\eta > \Delta
$.
The infinitesimal imaginary number is added to its argument in order to prevent
the branching points and the poles of the
above functions from entering the upper half plain of the complex
variable $\eta$.
The advanced Green functions can be obtained from the retarded one by a formal
substitution $\tau\rightarrow -\tau$ and $\ep_R(\eta)\rightarrow\ep_A(\eta)$
where
\be
	\ep_A(\eta) = \sqrt{(\eta- \im 0)^2 - \Delta^2}
\label{A-epsilon-advanced}
\ee
\par

According to the above consideration in Section 4 the component of the response
tensor
$Q_{zz}$ is given by the following expression
\be
     Q^{M} (\omega_n, \vec{k}) =
	- \frac{ \im e^2}{c} \int
     \frac{d^3p}{(2 \pi)^3}
     v_{z}^2(p_z)
	\Ga(\om_n,\xi_+,\xi_-)
\label{A-Q-matsubara}
\ee
where $v_z\,=\,(2\pi\ga/d) \sin (2\pi\ga p_z /d)$.
\par
Keeping the external frequency in the subset of discrete imaginary numbers:
$\om= \im\om_n = \im 2\pi n T$ we can transform the sum in the r.h.s. of
eq. (\ref{A-pseudo-vertex})
into an integral:
\bea
K^R (\im\om_n)\,=\,
\frac{1}{4\pi\im} \,\int_{\Ga_1}G^R(\eta +\im\om_n,\xi_+) G^R(\eta,\xi_-)
\tanh \left( \frac{\eta}{2T} \right)
d\eta +
\nonumber\\
\frac{1}{4\pi\im} \, \int_{\Ga_2}G^R(\eta +\im\om_n,\xi_+) G^A(\eta,\xi_-)
\tanh \left( \frac{\eta}{2T} \right)
d\,\eta +
\nonumber\\
\frac{1}{4\pi\im}\int_{\Ga_3}G^A(\eta +\im\om_n,\xi_+) G^A(\eta,\xi_-)
\tanh \left( \frac{\eta}{2T} \right)
d \,\eta
\qquad
\label{A-tanhtransform}
\eea
where the contours $\Ga_1,\Ga_2;\Ga_3$ in the complex plane $\eta$ go around
the intervals
of  the imaginary axis from $\eta =\im 0$ to $\eta =\im\infty$, from
$\eta =\,-\im\om_n$ to
$\eta =\im 0$ and from $\eta = -\im\infty$ to $\eta = -\im\om_n$
respectively, all counterclockwise.
These contour integrals can be readily transformed into the integrals along the
real axis,
using the periodicity of $\tanh(\eta /2T)$ with the period $\om_n$. The last
step is to put
$\om + \im 0$ instead of $\im\om_n$ in all integrals to get
the "retarded" analogue of the vertex (\ref{A-pseudo-vertex})
written in the following integral form
\bea
	\Ga^R(\om,\xi_+,\xi_-)
	 =
	\int_{-\infty}^{\infty}
	\frac{d \eta}{4 \pi \im}
	\left[
	\tanh \left(\frac{\eta_+}{2T} \right)
	-
	\tanh \left( \frac{\eta}{2T} \right)
	\right]
	\tilde{\Ga}^{RA}(\om,\eta ,\xi_+,\xi_-)
\nonumber\\
	\int_{-\infty}^{\infty}
	\frac{d \eta}{4 \pi \im}
	\tanh \left( \frac{\eta}{2T} \right)
	\left[
	\tilde{\Ga}^{RR}(\om,\eta ,\xi_+,\xi_-)
	-
	\tilde{\Ga}^{AA}(\om,\eta ,\xi_+,\xi_-)
	\right]
	\qquad
\label{A-pseudo-vertex-ret}
\eea
where double-retarded, double-advanced and mixed vertices are defined by the
following equations:
\bea
       \tilde{\Ga}^{RR}(\om,\eta ,\xi_+,\xi_-)
	 =
	G^R ( \eta_+, \xi_+)
	G^R ( \eta , \xi_-)
	-
	\qquad  \qquad  \qquad \qquad  \qquad \qquad
\nonumber\\
	G^R_0 ( \eta , \xi_+)
	G^R_0 ( \eta , \xi_-)
	+
	\bar{F}^R ( \eta_+, \xi_+)
	F^R ( \eta , \xi_-)
	\qquad  \quad
\label{A-ret-ret}\\
	\tilde{\Ga}^{AA}(\om,\eta ,\xi_+,\xi_-)
	 =
	G^A ( \eta , \xi_+)
	G^A ( \eta_-, \xi_-)
	-
	\qquad  \qquad  \qquad \qquad  \qquad \qquad
\nonumber\\
	G^A_0 ( \eta , \xi_+)
	G^A_0 ( \eta , \xi_-)
	+
	\bar{F}^A ( \eta , \xi_+)
	F^A ( \eta_-, \xi_-)
	\qquad \quad
\label{A-adv-adv}\\
	\tilde{\Ga}^{RA}(\om,\eta ,\xi_+,\xi_-)
	 =
	G^R ( \eta_+, \xi_+)
	G^A ( \eta ,\xi_-)
	+
	\bar{F}^R (\eta_+, \xi_+)
	F^A ( \eta , \xi_-)
	\qquad  \quad
\label{A-ret-adv}
\eea
Here and in what it follows $\eta_{\pm}= \eta \pm \om$.
The expressions (\ref{A-ret-ret}-\ref{A-ret-adv}) are obviously analytic
in the upper half-plane and since all the integrals are now convergent,
one can safely perform the integration of it over $ \xi$ first. This
integration becomes especially easy provided the representation
(\ref{A-retarded-green-functions}) and its analogue for advanced
Green functions are used.
The contribution due to the subtracted terms vanishes since
\bea
	G^R_0 (\eta , \xi) =
	\frac{1 }{ \eta + \im /2 \tau - \xi}
\nonumber\\
	G^A_0 (\eta , \xi) =
	\frac{1 }{ \eta - \im /2 \tau - \xi}
\label{A-green-normal}
\eea
The result of integration of vertex
$\Ga^R (\om,\xi_+,\xi_-)$
(see eq. (\ref{A-pseudo-vertex-ret})) over $\xi$
depends on the scalar product $\vec{k}\vec{v}$. The remaining
integration over $\phi$ and $p_z$ becomes trivial, once
the spatial dispersion is neglected, leading to the following final
expression:
\bea
	K^R (\om) =
	\int_{-\infty}^{\infty}
	\frac{d \eta}{4}\left\{
	\left(
	1 +
		\frac{\eta_+ \eta + \Delta^2
				}{\ep_R (\eta_+) \ep_A (\eta)}
	\right)
	\frac{\tanh(\eta_+/2T)
	-
	\tanh(\eta/2T)
		}{\ep_R (\eta_+) - \ep_A (\eta) + \im / \tau }
	\right. \nonumber\\
	+
	\left.
	\tanh \left( \frac{\eta}{2T} \right) \left[
	\left(1 -
	\frac{\eta \eta_+ + \Delta^2}{\ep_R (\eta) \ep_R (\eta_+)}
	\right)
	\left(\ep_R (\eta) + \ep_R (\eta_+) + \im / \tau \right)^{-1}
	\right. \right.
\nonumber\\
       + \left. \left.
	\left(1 -
		\frac{\eta \eta_- +
			\Delta^2}{\ep_A(\eta) \ep_A(\eta_-)}
	\right)
	\left(\ep_A (\eta) + \ep_A (\eta_-) - \im / \tau\right)^{-1}
	\right] \right\}
	\qquad
\label{A-after-xi-integration}
\eea
The real and imaginary parts of $K^R(\om)$ for $\om < 2\De$  are given by eqs.
(\ref{ReK},\ref{ImK}) of the text. See Appendix B for the expressions of
the same values and their analysis in the range $\om > 2\De$.

\renewcommand{\thesection}{Appendix \Alph{section}}
\setcounter{section}{1}
\section{Above the Threshold.}
\label{app:above-threshold}

\renewcommand{\thesection}{\Alph{section}}
\renewcommand{\theequation}{\thesection.\arabic{equation}}
\setcounter{equation}{0}

Here we present formal expressions for the real and imaginary parts of the
normalized Pippard kernel in the range $\om > 2\De$ and a brief analysis of
their behavior at zero temperature.
The required expressions can be extracted straightforwardly
from the general equation (\ref{A-after-xi-integration}):
\bea
{\cal R}e K(\om)\,= \,
\int_{-\De}^{\De}
\frac{d\eta}{\ep_{+}^2 + (\bar{\varepsilon} + 1/\tau )^2}
\left[
\ep_{+}\,+\,
\left(\bar{\varepsilon} + \frac{1}{\tau}\right)
\frac{\eta_+\eta + \De^2}{\ep_+\bar{\varepsilon}}
\right]
\tanh\frac{\eta_+}{2T}
+
\nonumber\\
\int_{\De}^{\infty}\left[\frac{\ep_{+}+\ep}{(\ep_{+}+\ep)^2
+ 1/\tau^2}
\left( 1\,-\,\frac{\eta_+\eta + \De^2}{\ep_+\ep}\right)
\left(\tanh\frac{\eta_+}{2T}\,+\,
\tanh\frac{\eta}{2T}\right) \right.
+
\nonumber\\
\left.
\frac{\ep_{+}-\ep}{(\ep_{+}-\ep)^2 + 1/\tau^2}
\left( 1\,+\,\frac{\eta_+\eta + \De^2}{\ep_+\ep}\right)
\left(\tanh\frac{\eta_+}{2T}\,-\,
\tanh\frac{\eta}{2T}\right)\right] \frac{d\eta}{2} \,
+
\nonumber\\
\int_{\De }^{\om - \De } \frac{d\eta}{2}
\left[
\frac{\ep-\ep_{-}}{(\ep-\ep_{-})^2 + 1/\tau^2}
\left( 1\,+\,\frac{\eta_-\eta + \De^2}{\ep_+\ep}\right)
\tanh\frac{\eta}{2T}
\right.
+
\nonumber\\
\left.
\frac{1}{2} \frac{\ep+\ep_{-}}{(\ep+\ep_{-})^2 + 1/\tau^2}
\left( 1\,-\,\frac{\eta\eta_{-} + \De^2}{\ep \ep_{-}}\right)
\left(\tanh\frac{\eta}{2T}\,-\,
\tanh\frac{\eta_{-}}{2T}\right)
\right]
\qquad
\label{A-rek}
\eea
\bea
{\cal I}m K(\om)\,=\frac{1}{2\tau}
\int_{\De }^{\om - \De } d\eta
\left[
\frac{\tanh(\eta/2T)}
{(\ep-\ep_{-})^2 + 1/\tau^2}
\left( 1\,+\,\frac{\eta_-\eta + \De^2}{\ep_-\ep}\right)
\right.  \qquad
-
\nonumber\\
\left.
\frac{1}{2} \frac{1}{(\ep+\ep_{-})^2 + 1/\tau^2}
\left( 1\,-\,\frac{\eta\eta_{-} + \De^2}{\ep \ep_-}\right)
\left(\tanh\frac{\eta}{2T}\,-\,
\tanh\frac{\eta_{-}}{2T}\right)\right]
+
\nonumber\\
\frac{1}{2\tau}
\int_{\De}^{\infty} d\eta
\left[
\left( 1\,-\,\frac{\eta_+\eta + \De^2}{\ep_+\ep}\right)
\left((\ep_{+}+\ep)^2 + \frac{1}{\tau^2} \right)^{-1}
\right.
-
\nonumber\\
\left.
\left( 1\,+\,\frac{\eta_+\eta + \De^2}{\ep_+\ep}\right)
\left( (\ep_{+}-\ep)^2 + \frac{1}{\tau^2} \right)^{-1}
\right]
\left(\tanh\frac{\eta_+}{2T}\,-\,\tanh\frac{\eta}{2T} \right)
\qquad
\label{A-imk}
\eea
The notations are the same as in the text.
\par
\label{pg:normal-limit}
Let us first take a look at the normal limit
$ \De \longrightarrow 0 $.
In this case the first integral in the l.h.s. of the eq. (\ref{A-rek})
just vanishes whereas the coherence factors
$(1 \pm (\eta \, \eta_{\pm} + \De^2)/ \ep \ep_{\pm})$
are reduced either to 0 or to 2
due to the following identities:
$\ep_+ = \eta_+$, $\ep = \eta$ and $\ep_- = - \eta_-$.
The Lorentzian factors associated with the non-vanishing
coherence factors tend to
$\om/(\om^2 + 1/\tau^2)$ or $(1/\tau)/(\om^2 + 1/\tau^2)$
in formulas for the real and the imaginary parts of
the kernel respectively. Taking into account
the above remarks one can evaluate the remaining integrals
straightforwardly and finally arrive to the Drude-Lorentz formula
(\ref{eq:normal-limit}).

We specially consider the case of zero temperature.
Then equations (\ref{A-rek}) and (\ref{A-imk}) look much simpler:
\bea
{\cal R}e K(\om)\, =
\int_\De^{\infty}\frac{\ep_{+}+\ep}{(\ep_{+}+\ep)^2 + 1/\tau^2}
\left( 1-\frac{\eta_+\eta + \De^2}{\ep_+\ep}\right)d\eta
+
\nonumber\\
\half \int_{\De }^{\om - \De } \,  d\eta
\left[
\frac{\ep-\ep_{-}}{(\ep-\ep_{-})^2 + 1/\tau^2}
\left( 1\,+\,\frac{\eta_-\eta + \De^2}{\ep_-\ep}\right)
\right. +  
\nonumber\\
\left.
\frac{\ep+\ep_{+}}{(\ep+\ep_{+})^2 + 1/\tau^2}
\left( 1\,-\,\frac{\eta_+\eta + \De^2}{\ep_+\ep}\right)
\right] \,+
\nonumber\\
\int_{-\De}^{\De}
\left[\ep_{+}\,+
\left(\bar{\varepsilon} + \frac{1}{\tau}\right)
\frac{\eta_+\eta + \De^2}{ \ep_+\bar{\varepsilon}}
\right]
\frac{d\eta}{
\left(\ep_{+}^2
+ (\bar{\varepsilon} + 1/\tau )^2 \right)}
\label{A-rek0}
\eea
\bea
{\cal I}m K(\om)\,=
\frac{1}{2\tau}
\int_{\De }^{\om - \De }
\left[
\left( 1\,+\,\frac{\eta_- \eta + \De^2}{\ep_-\ep}\right)
\left( (\ep-\ep_{-})^2 + \frac{1}{\tau^2} \right)^{-1}
\right. -
\nonumber\\
\left.
\left( 1\,-\,\frac{\eta_- \eta + \De^2}{\ep_-\ep}\right)
\left( (\ep+\ep_{-})^2 + \frac{1}{\tau^2} \right)^{-1}
\right]
d\eta
\label{A-imk0}
\eea
The imaginary part of (\ref{A-imk0}) corresponds to the absorption of a photon
accompanied
by the creation of two quasiparticles (dissociation of a Cooper pair).  The
probability of this
process, suppressed in the absence of scatterers, is proportional to the
scattering rate
$1/\tau$. In addition to the coherence factor it contains the Lorentzian
distribution  function for
the total momentum of a pair of quasiparticles located on the opposite points
of a diameter. When $\tau \rightarrow\infty$,
this distribution turns into $\de$-function.
\par
The  integral in the above formula (\ref{A-imk0}) for the imaginary part may be
expressed in terms
of  the standard elliptic integrals as follows:
\begin{eqnarray}
	{\cal I}m K^R(w) =
	 {{{4\,{{\Delta}^2}\,\tau}
	\,{\Pi(n(\omega ,\tau ),m(\tau ))}}
	\over {\left( \omega + 2\,\Delta  \right) \,
       \left( 1 - 4\,{{\Delta}^2}\,{{\tau }^2} +
	 {{\omega }^2}\,{{\tau }^2} \right) }}
	\,\left( 1 -
	 \frac{(2\,\Delta\,\omega \,{{\tau }^2)^2} }{
	  \left(1 + {{\omega }^2}\,{{\tau }^2} \right)^2}
	  \right)
\nonumber\\
	-{{\left( \omega + 2\,\Delta  \right) \,\tau
	}\over {1 + {{\omega }^2}\,{{\tau }^2}}} \, E(m(\omega )) +
   {{4\,\Delta\,\tau \,\left( \Delta + \omega  +
	 {{\omega }^3}\,{{\tau }^2} \right) }\over
     {\left( \omega + 2\,\Delta  \right) \,
       {{\left( 1 + {{\omega }^2}\,{{\tau }^2} \right) }^2}}}
       \,K(m(\omega)) \qquad
\label{A-elliptic}
\end{eqnarray}
Here ${\rm K}\left(m\right)$, ${\rm E}\left(m\right)$ and
${\rm \Pi}\left(n, m\right) $ are complete elliptic integrals of the first,
the second and the third kind respectively. The moduli of the integrals
depend on the frequency and the scattering rate.
\begin{equation}
  m(\omega) = \left({{\omega  - 2 \Delta } \over { \omega + 2 \Delta }}
	\right)^2,
  \qquad
  n(\omega,\tau) = -{{{ {\tau^2} \, {\left( \omega - 2 \Delta \right) }^2}\,
	    \left( 1 + {\omega^2} \, {\tau^2} \right) }\over
	  {\left(1 +  \left({\omega^2} -4 {\Delta^2} \right) \, {\tau^2}
	  \right) }}
\label {A-moduli}
\end{equation}
The  function (\ref{A-elliptic}) is always negative.
Just above the threshold its absolute value
increases linearly with the intercept being inversely proportional
to the scattering rate:
\begin{equation}
	{\cal I}m K(\omega) =
	-{{\pi \, \tau \left( \omega - 2 \Delta \right) }}/2 +
	{{{\rm O}(\omega/\Delta - 2)}^2}
\label{A-linear}
\end{equation}
The non-linear phase of the descending slope may be approximated
by means of the  following interpolating function:
\begin{equation}
{\cal I}m K(\omega)  \approx
{{\pi } \over {4\, \Delta \, \tau}} \,
\left({{1} \over {\sqrt{1 +
4\,\left( \omega - 2 \Delta \right)\, \Delta \tau^2 }}} - 1 \right)
\label{A-imk0-interpol}
\end{equation}
This interpolation gives a remarkably good estimate of the actual
${\cal I}m K(\omega)$ almost up to the point
$\om_{min} \,  \approx \, 2\,\Delta + 0.9/  \tau$
where the imaginary part reaches its minimum
\footnote{The approximate dependence of the minimum position
on a scattering  rate has been found numerically.}.
A minimal value of ${\cal I}m K(\omega)$  significantly
depends on the scattering rate. If the latter is small
compared to the threshold frequency
($\Delta \, \tau \gg 1$) the dissipation is also small:
${\cal I}m K(\omega_{min}) \, \approx \, -\pi/4\,\Delta\,\tau$.
However, in the regime of a strong scattering the
dissipation growth is saturated at the level:
\be
	\lim_{\Delta \, \tau \rightarrow 0}{
	{\cal I}m K(\omega_{min}) = -\half}
\label{A-imk0-min}
\ee
Examining eq. (\ref{A-imk0-interpol})
one can easily see the appearance of the  intermediate frequency scale
scale $1/ \De \tau^2$.

In the high frequency range the imaginary part of the Pippard
kernel decays monotonically with the asymptotics:
\be
	{\cal I}m K(\om )\,= \, - 1/ \om \, \tau  + {\rm  O}(1/\om^2)
\label{A-imk0-high-freq}
\ee
in conformity with the Drude-Lorentz
formula for the normal metal (\ref{eq:normal-limit}).
The existence of the gap becomes irrelevant for the plasma
oscillations driven by the electromagnetic wave provided the
frequency of the latter is large compared to the threshold
value.
\par
The discontinuity of the derivative with respect to frequency of the
imaginary part of the Pippard  kernel  at $\om = 2\, \De$
implies that the total Pippard  kernel, considered as a function of the
complex variable $\om$, has a logarithmic-type singularity
at this point.
The behavior of this function in the vicinity of its singularity is governed
by the eq. (\ref{eq:log-sing}).
The real part of the kernel decreases rapidly when approaching the
threshold with a negative logarithmically divergent derivative.
\label{pg:log-sing}
\par
Switching the  temperature on causes the augmentation
of the dissipation. On the contrary,  ${\cal R}e K(\om )$
at finite temperature is diminished.
Still near the threshold the asymptotic behavior
(\ref{A-imk0-interpol})
is valid if one considers this contribution as complementary to the
Cherenkov absorption part, and multiplies it by the factor
$\tanh(\De/2T)$.

\newpage
\begin{center}
	{\Large \bf Figure Captions.}
\end{center}

\vspace{0.3in}
\noindent
{\large \bf Figure 1}

\vspace{0.1in}
\noindent
Reflectivity versus frequency (measured in relative units:
$ \omega / \Om_z $ ) in the normal state.
\begin{description}
\item[Figure 1A:]
The high-frequency dielectric constant $\E = 25$.
\item[Figure 1B:]
The high-frequency dielectric constant $\E = 4$.
\end{description}
\noindent
The different curves on both plots correspond to the different values
of scattering  rate $\Omega_z \tau =  0.01;\, 0.1;\,  0.5;\,  1.0;\,5.0$.

\vspace{0.2in}
\noindent
{\large \bf Figure 2}

\vspace{0.1in}
\noindent
Real part of the normalized Pippard kernel versus frequency
(measured in relative units:
$ \omega /  T_c $ ). The bare plasma frequency  $\Om_z =1.2\, T_c$.
The high-frequency dielectric constant $\E = 25$ .
Three spectra are plotted  for each of the following scattering rates:
\begin{description}
\item[Figure 2A:]
$\tau  T_c=0.1$.
\item[Figure 2B:]
 $\tau T_c=3.4$
\end{description}
The dashed plot corresponds to
$T/T_c=0.30$,
$\De/T_c=1.73$
and the dashed-dotted one corresponds to
$T/T_c=0.74$ ,
$\De/T_c=1.38$
respectively. The values of $\De(T/T_c)$ have been calculated
in the frameworks of the BCS theory.

\vspace{0.2in}
\noindent
{ \large \bf Figure 3}

\vspace{0.1in}
\noindent
Imaginary part of the normalized Pippard kernel.
The bare plasma frequency  $\Om_z =1.2\, T_c$.
The high-frequency dielectric constant $\E = 25$ ,
The dashed plot corresponds to
$T/T_c=0.30$,
$\De/T_c=1.73$
and the dashed-dotted one corresponds to
$T/T_c=0.74$ ,
$\De/T_c=1.38$.
\begin{description}
\item[Figure 3A:]
$\tau  T_c=0.1$.
\item[Figure 3B:]
 $\tau T_c=3.4$
\end{description}

\vspace{0.2in}
\noindent
{ \large \bf Figure 4}

\vspace{0.1in}
\noindent
Reflectivity versus frequency (measured in relative units:
$ \omega /  T_c $ ).  The bare plasma frequency  $\Om_z =1.2\, T_c$.
High-frequency dielectric constant $\E = 25$ . Three spectra are plotted  for
each of the following scattering rates:
\begin{description}
\item[Figure 4A:]
$\tau  T_c=0.1$.
\item[Figure 4B:]
 $\tau T_c=0.9$
\item[Figure 4C:]
$\tau T_c=1.7$
\item[Figure 4D:]
$\tau T_c=3.4$
\end{description}
The solid curve is a plot for the normal state.
The dashed plot corresponds to
$T/T_c=0.30$,
$\De/T_c=1.73$
and  the dashed-dotted one corresponds to
$T/T_c=0.74$ ,
$\De/T_c=1.38$
respectively.

\vspace{0.2in}
\noindent
{ \large \bf Figure 5}
\vspace{0.1in}
\noindent
Comparison of the imaginary parts of the normalized Pippard kernel
for the BCS and the two-fluid models. The solid curves correspond to the BCS
model.  The dashed plots correspond to the two-fluid model.
\newline
The bare plasma frequency  $\Om_z =1.2\, T_c$.
High-frequency dielectric constant $\E = 25$ ,
$T/T_c=0.30$,
$\De/T_c=1.73$ .
\begin{description}
\item[Figure 5A:]
$\tau  T_c=0.1$.
\item[Figure 5B:]
$\tau T_c=3.4$
\end{description}

\begin{figure}[b]
\PSbox{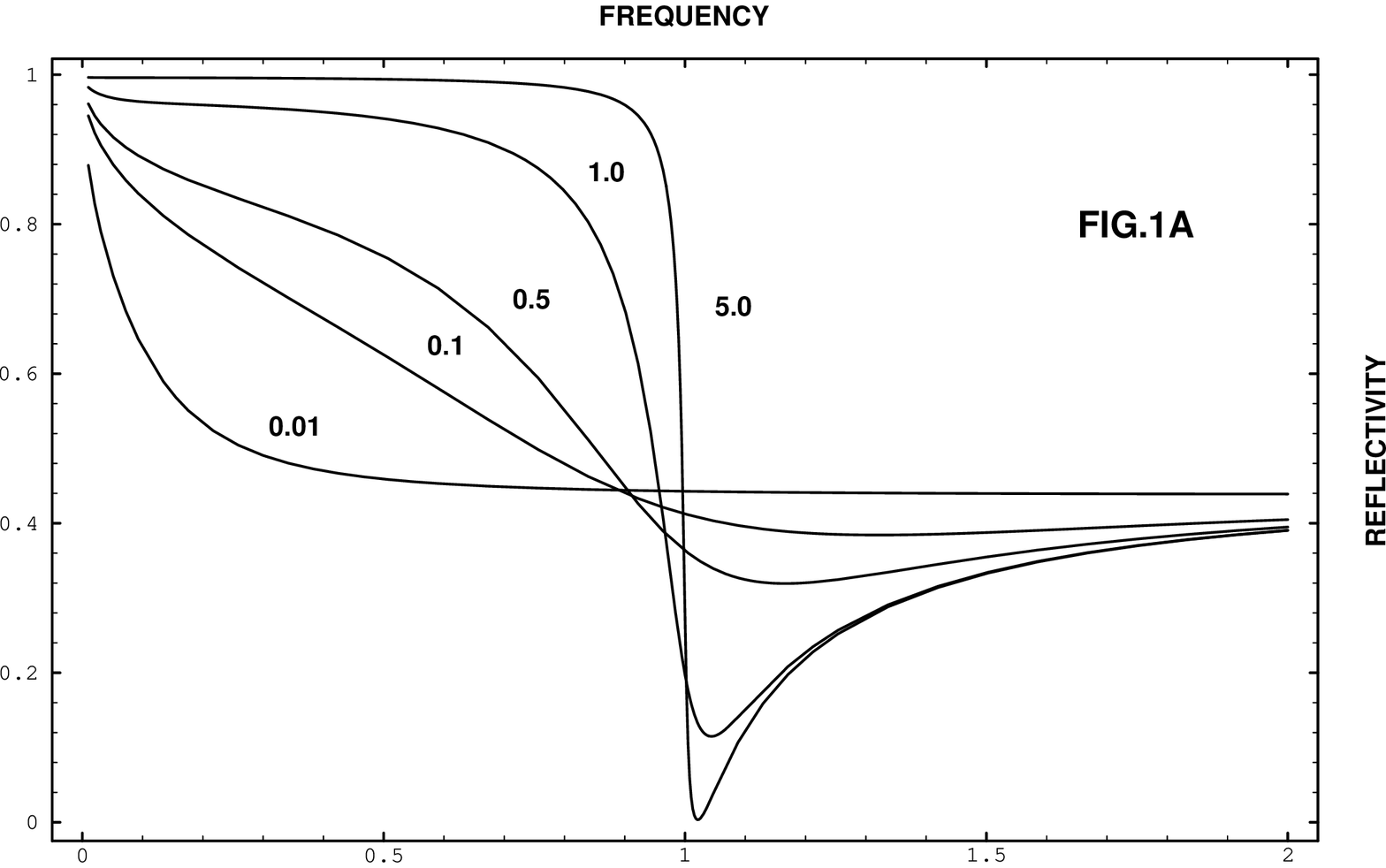 hoffset=0 voffset=-134 hscale=65 vscale=60}{5.3in}{3.5in}
\end{figure}
\begin{figure}[b]
\PSbox{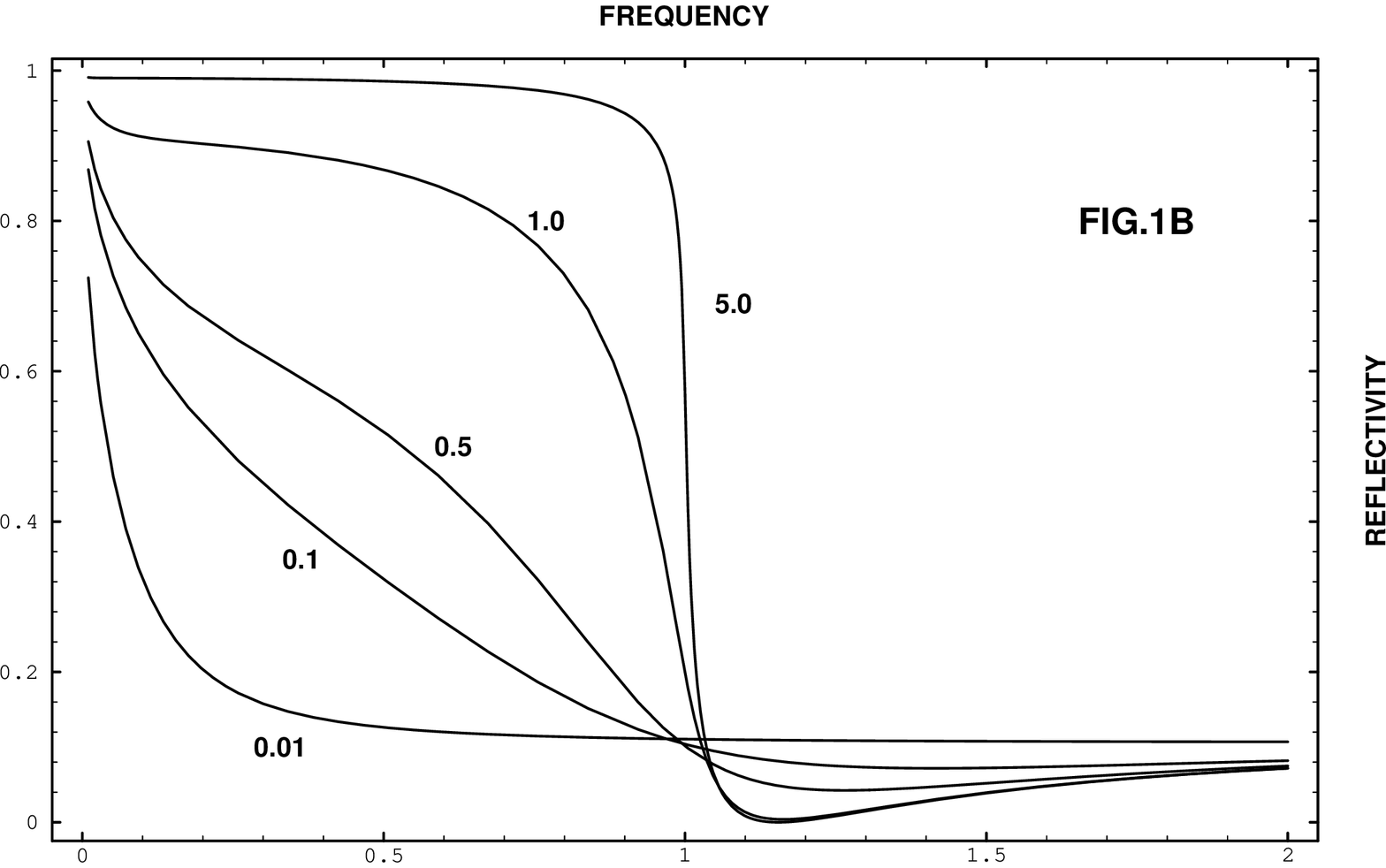 hoffset=0 voffset=-134 hscale=65  vscale=60}{5.3in}{3.5in}
\end{figure}
\begin{figure}[b]
\PSbox{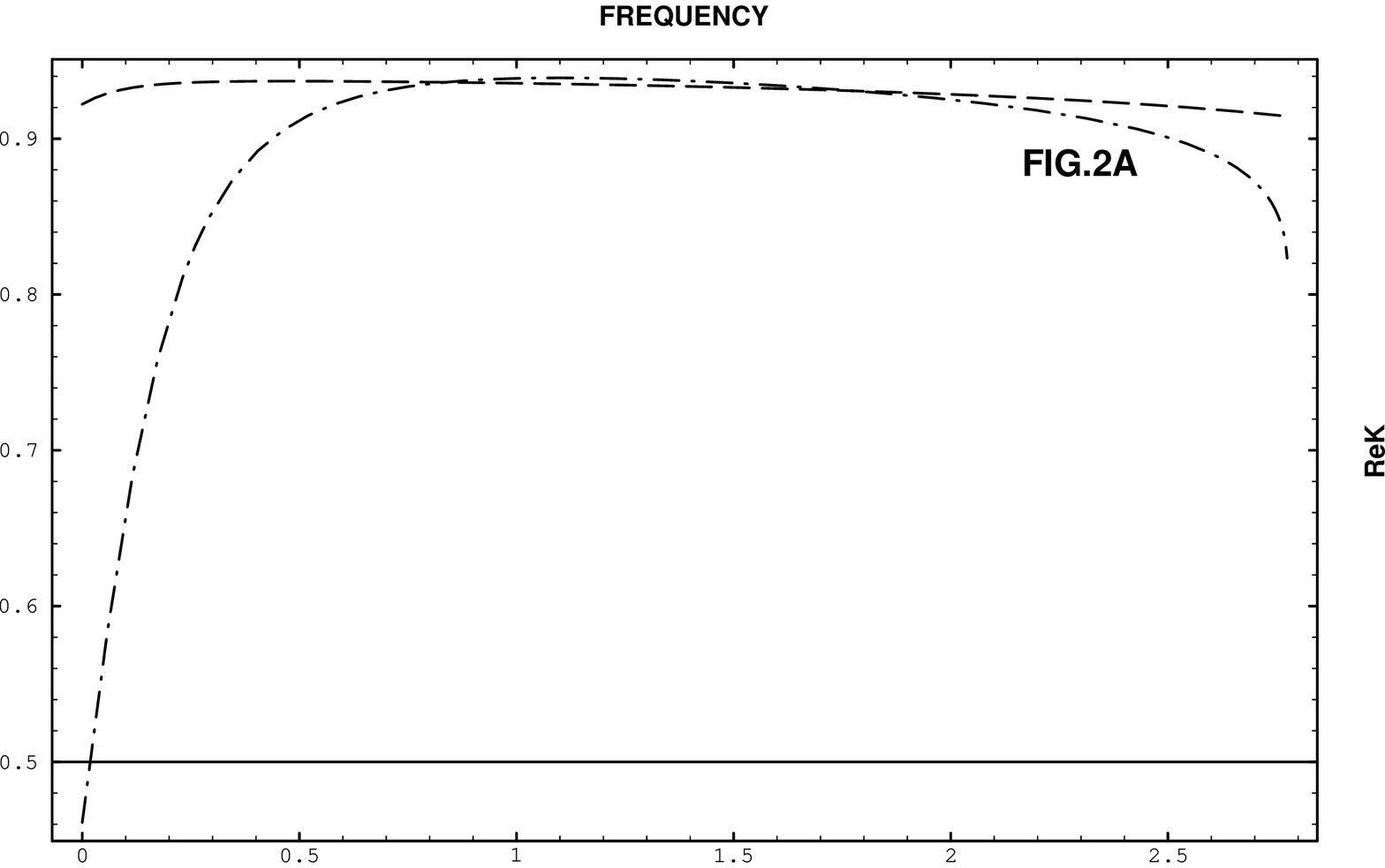 hoffset=0 voffset=-134 hscale=65 vscale=60}{5.3in}{3.5in}
\end{figure}
\begin{figure}[b]
\PSbox{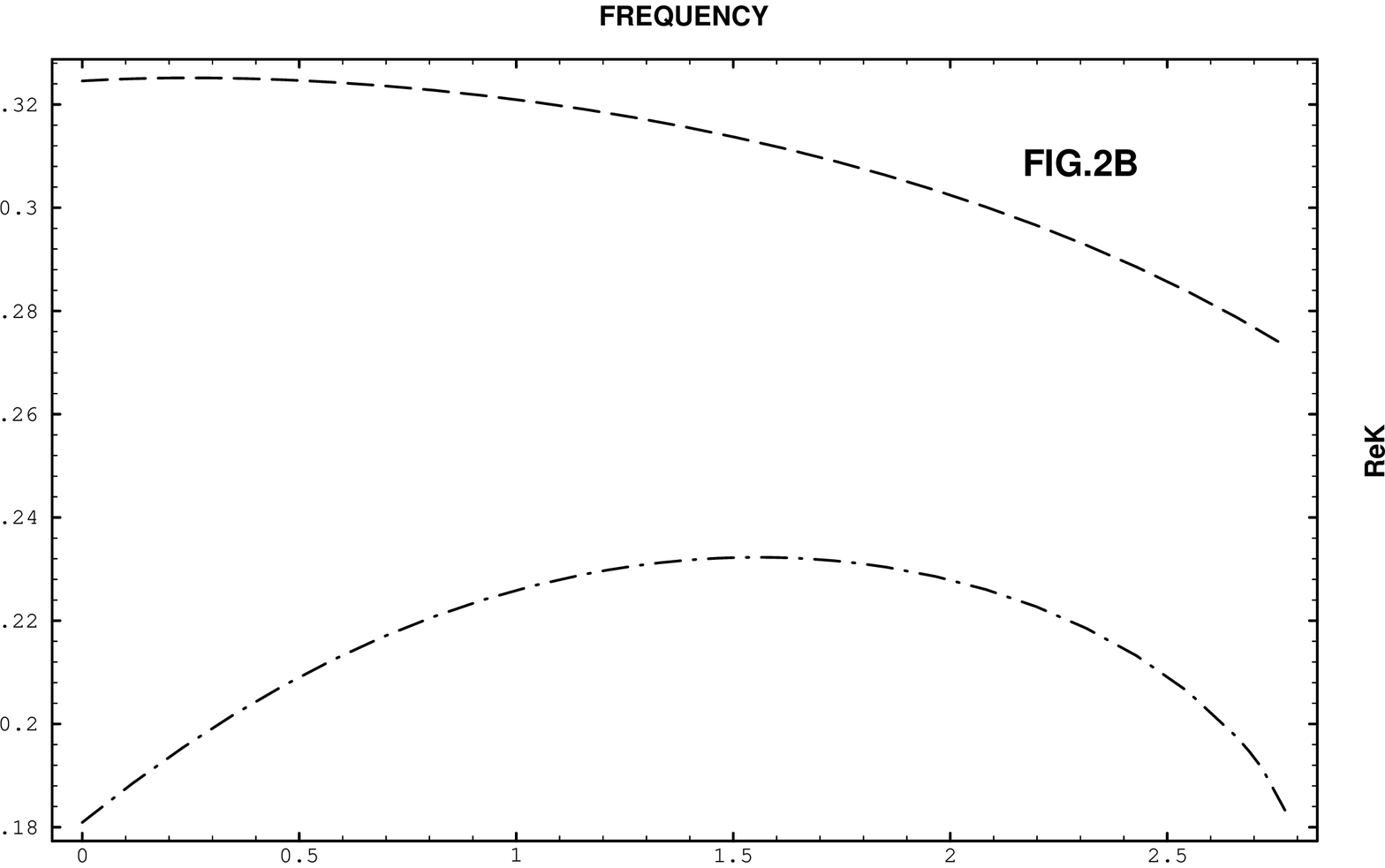 hoffset=0 voffset=-134 hscale=65 vscale=60}{5.3in}{3.5in}
\end{figure}
\begin{figure}[b]
\PSbox{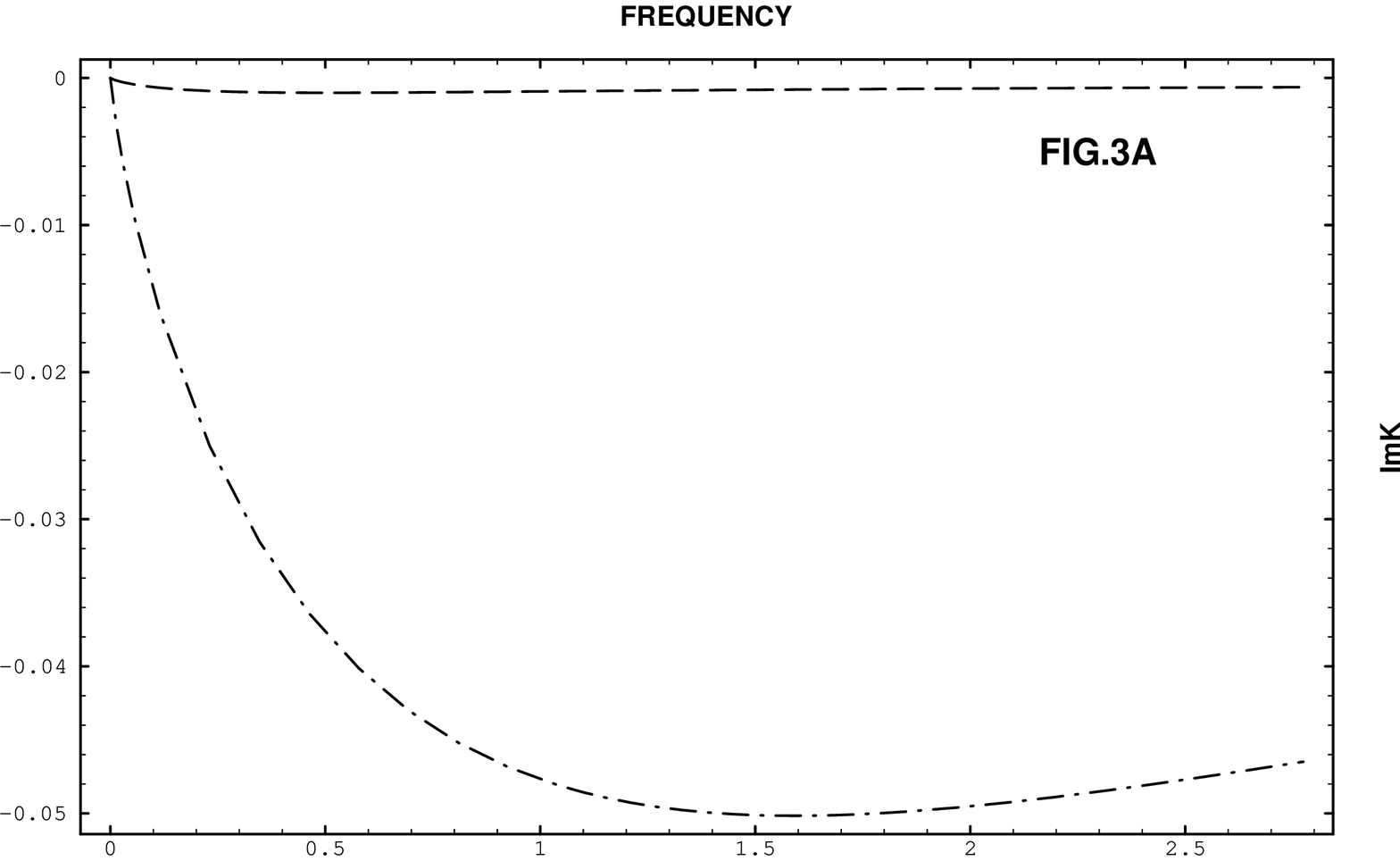 hoffset=0 voffset=-134 hscale=65 vscale=60}{5.3in}{3.5in}
\end{figure}
\begin{figure}[b]
\PSbox{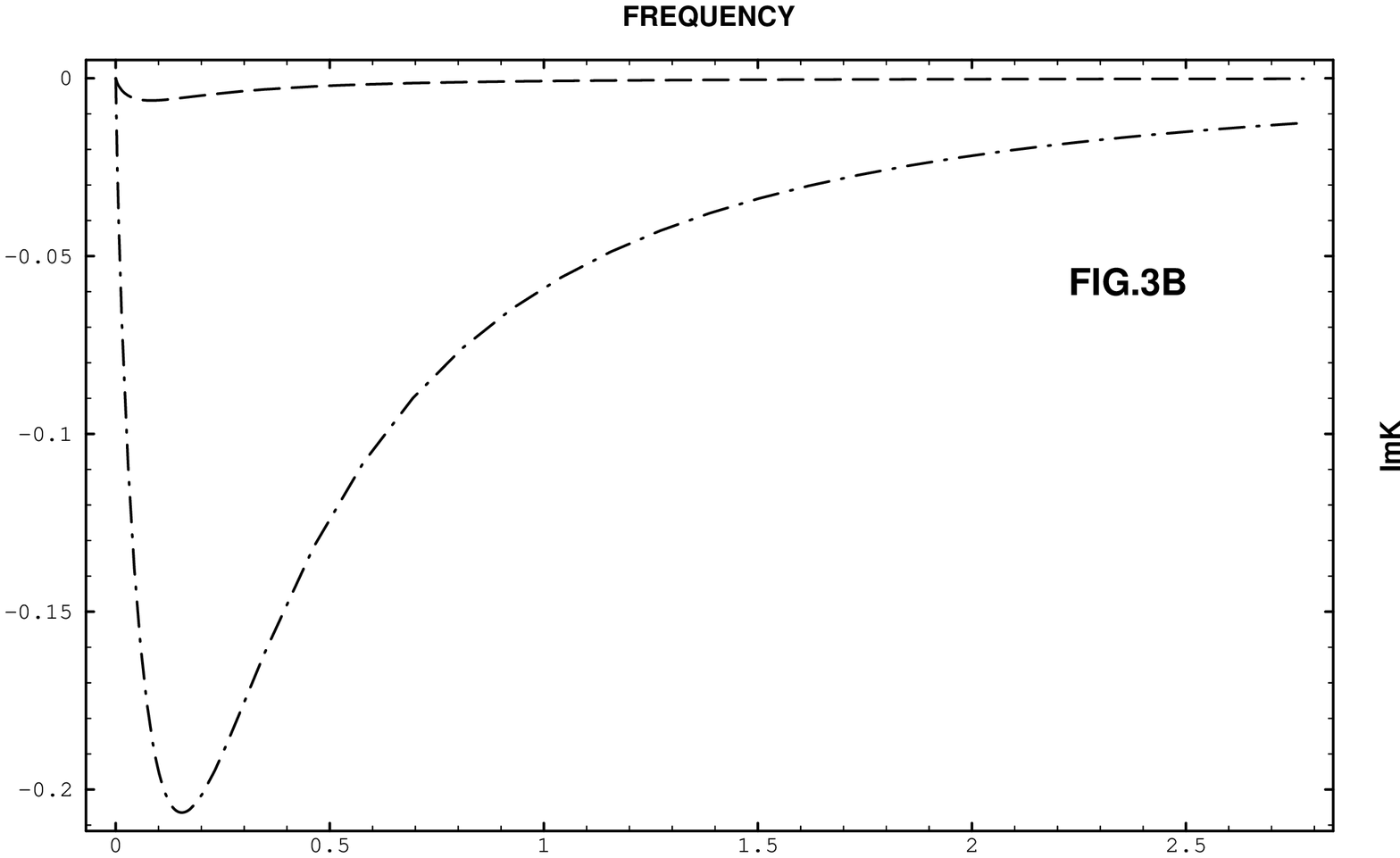 hoffset=0 voffset=-134 hscale=65 vscale=60}{5.3in}{3.5in}
\end{figure}
\begin{figure}[b]
\PSbox{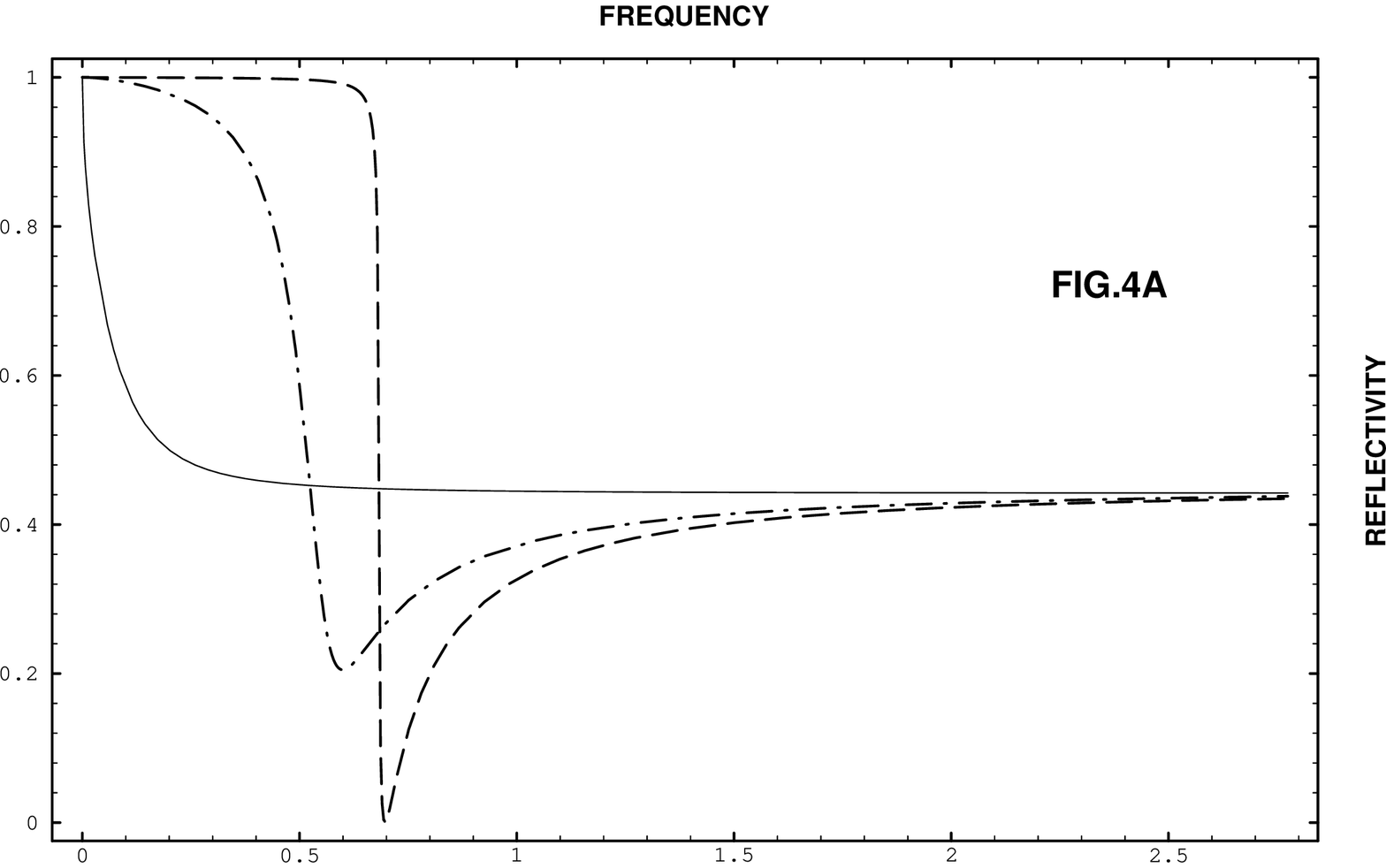 hoffset=0 voffset=-134 hscale=65 vscale=60}{5.3in}{3.5in}
\end{figure}
\begin{figure}[b]
\PSbox{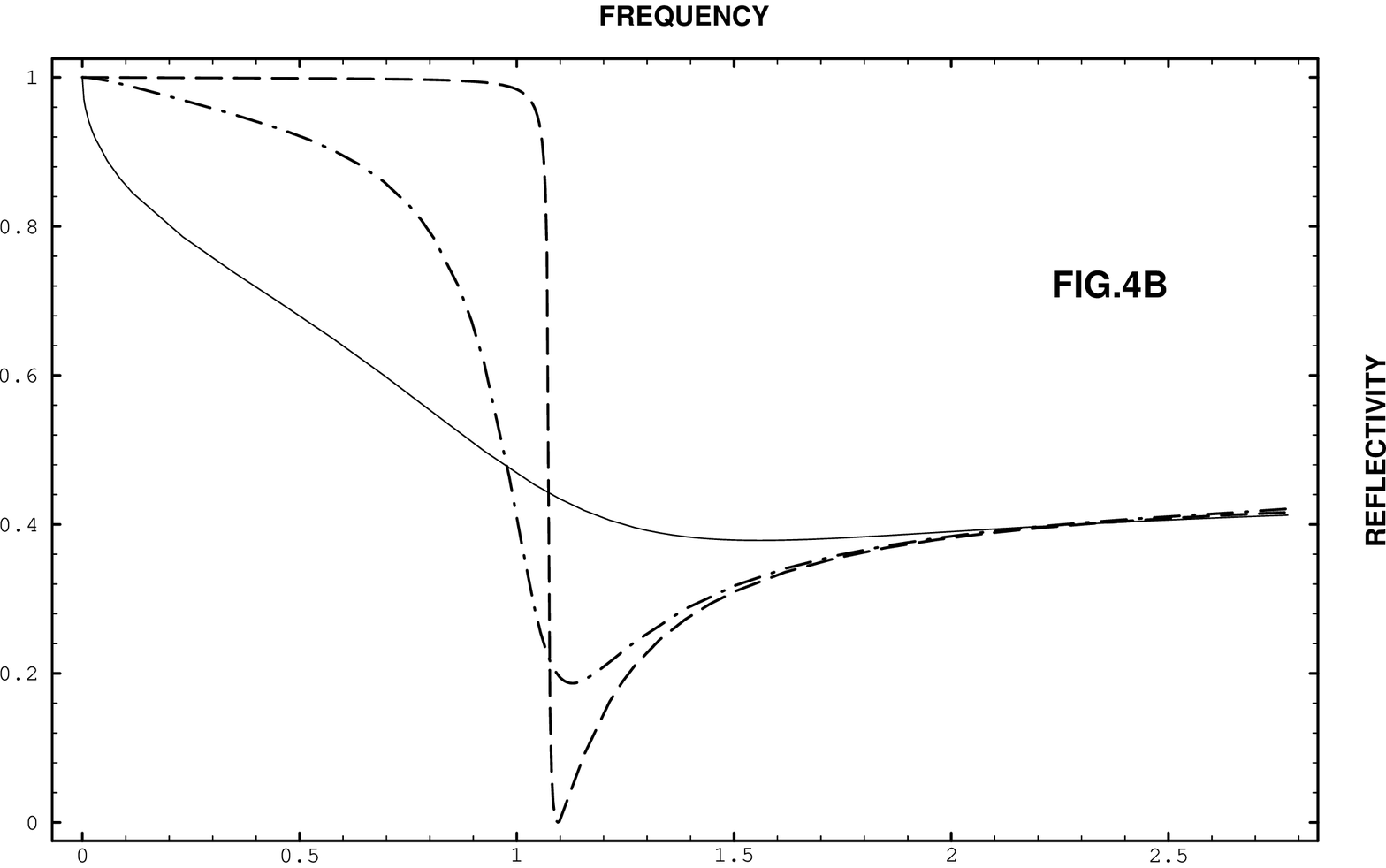 hoffset=0 voffset=-134 hscale=65 vscale=60}{5.3in}{3.5in}
\end{figure}
\begin{figure}[b]
\PSbox{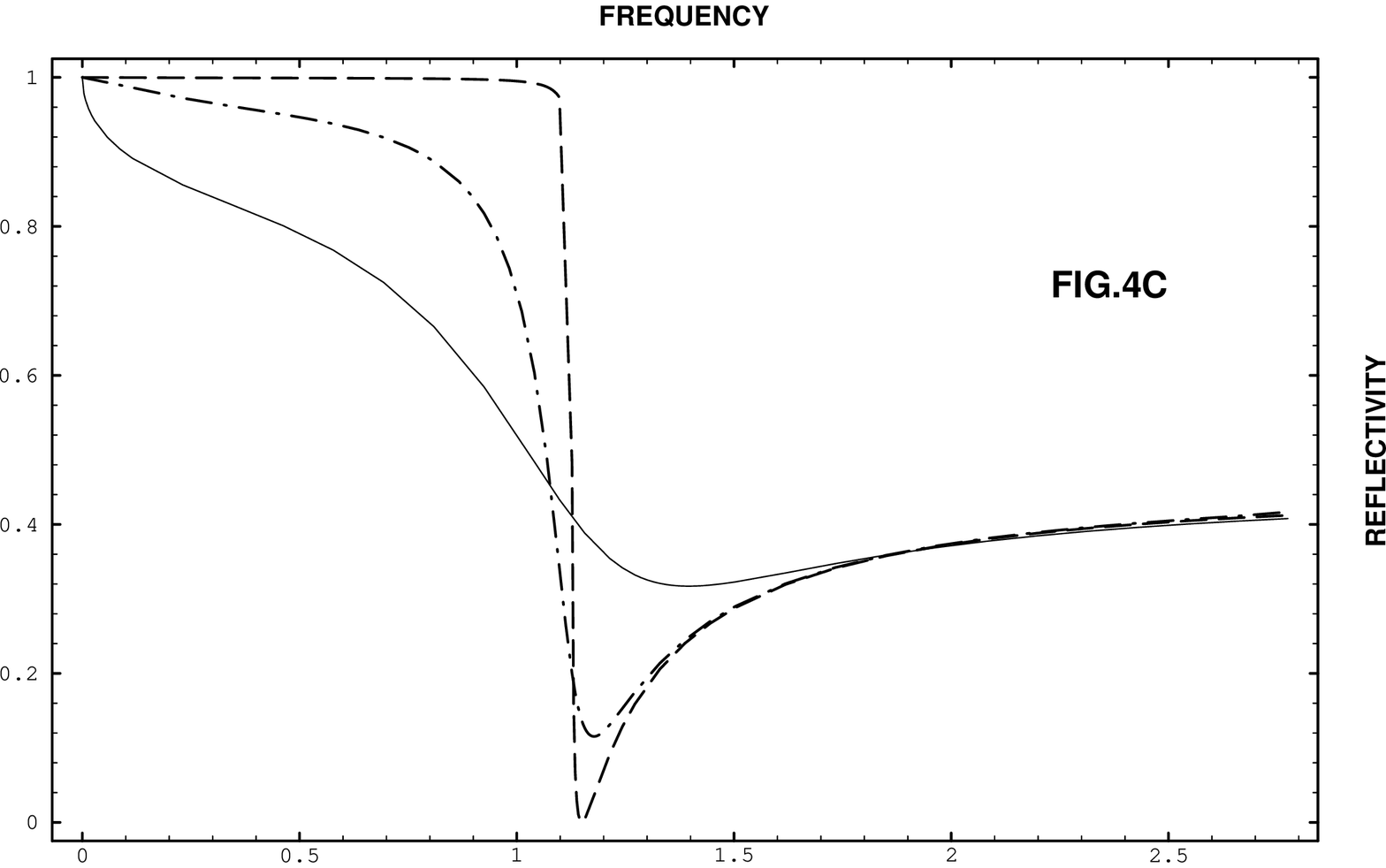 hoffset=0 voffset=-134 hscale=65 vscale=60}{5.3in}{3.5in}
\end{figure}
\begin{figure}[b]
\PSbox{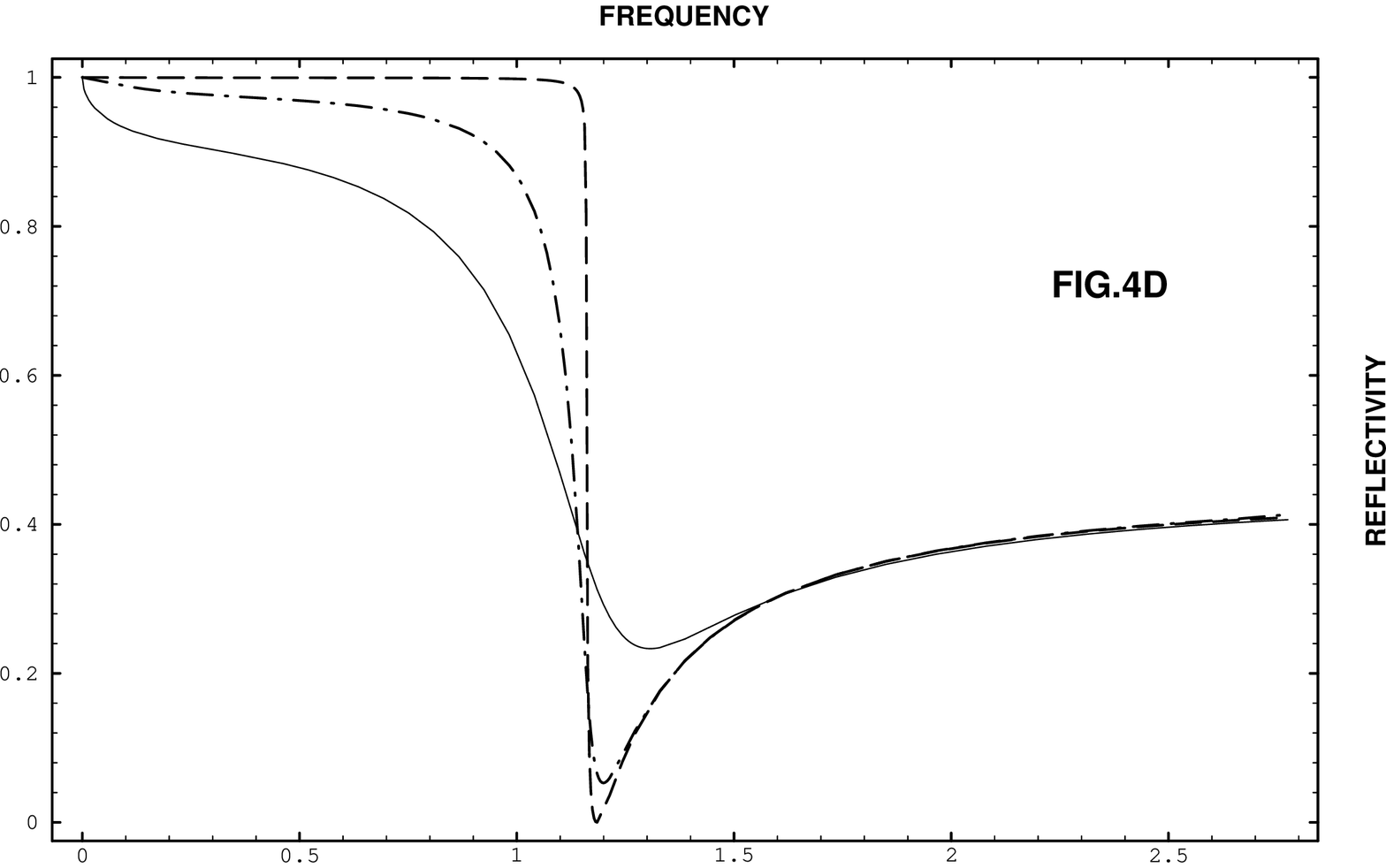 hoffset=0 voffset=-134 hscale=65 vscale=60}{5.3in}{3.5in}
\end{figure}
\begin{figure}[b]
\PSbox{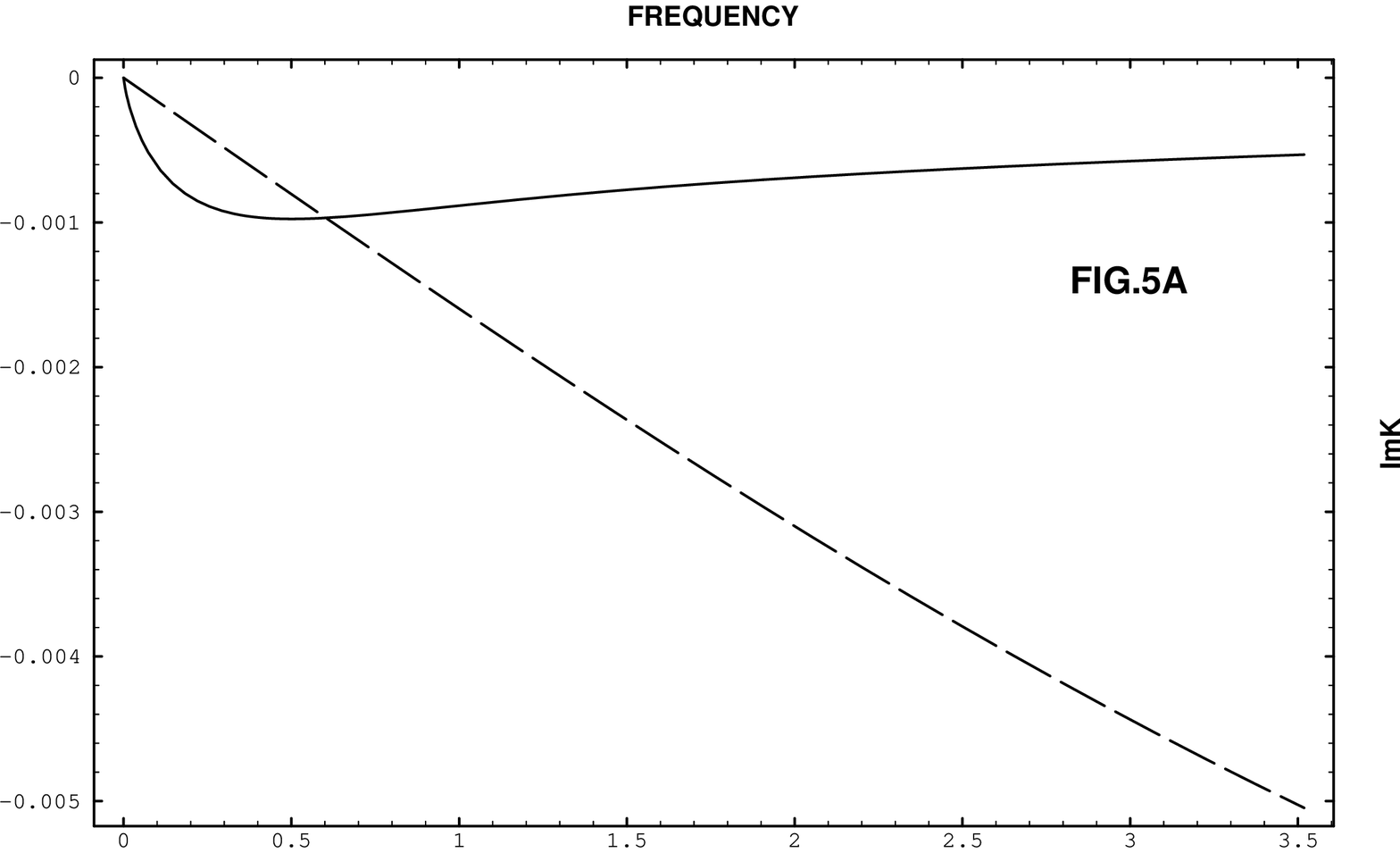 hoffset=0 voffset=-134 hscale=65 vscale=60}{5.3in}{3.5in}
\end{figure}
\begin{figure}[b]
\PSbox {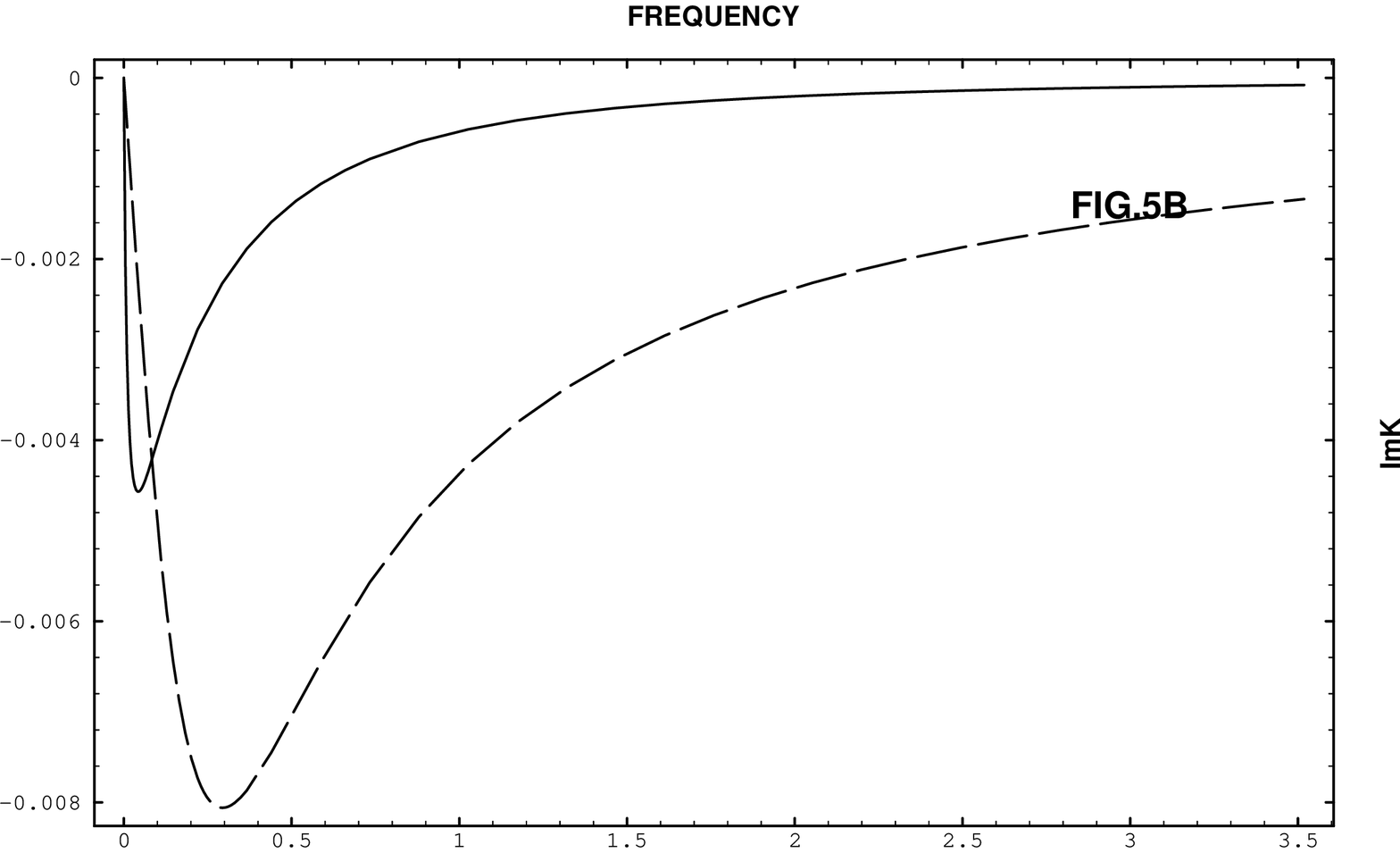 hoffset=0 voffset=-134 hscale=65 vscale=60}{5.3in}{3.5in}
\end{figure}

\end{document}